\documentclass[11pt,a4]{article}

\usepackage{rotating}
\usepackage{citesort}

\newcommand{\smf}[2]{\hspace{-1.8mm}\begin{array}{c} \\[-5.5mm] \frac{#1}{#2}\end{array}\hspace{-1.8mm}}
\newcommand{\hf}{\smf{1}{2}}
\newcommand{\roemII}{I\hspace{-0.5mm}I}
\newcommand{\mbf}{\mathbf}

\begin{document}

\title{\bf {The helium atom in a strong magnetic field}}

\author{W.Becken, P.Schmelcher and F.K. Diakonos\footnote{Permanent address: Department of Physics, 
Univ. of Athens, GR-15771 Athens, Greece}\\
Theoretische Chemie\\
Physikalisch-Chemisches Institut\\
Im Neuenheimer Feld 253\\69120 Heidelberg\\Federal Republic of Germany}

\date{}
\maketitle

\begin{abstract}
We investigate the electronic structure of the helium atom in a magnetic field between $B=0$ and 
$B=100a.u.$. The atom is treated as a nonrelativistic system with two interacting electrons and a 
fixed nucleus. Scaling laws are provided connecting the fixed-nucleus Hamiltonian to the one for 
the case of finite nuclear mass. Respecting the symmetries of the electronic Hamiltonian in the 
presence of a magnetic field, we represent this Hamiltonian as a matrix with respect to a two-particle 
basis composed of one-particle states of a Gaussian basis set. The corresponding generalized eigenvalue 
problem is solved numerically, providing in the present paper results for vanishing magnetic quantum 
number $M=0$ and even or odd $z$-parity, each for both singlet and triplet spin symmetry. Total 
electronic energies of the ground state and the first few excitations in each subspace as well as their 
one-electron ionization energies are presented as a function of the magnetic field, and their behaviour 
is discussed. Energy values for electromagnetic transitions within the $M=0$ subspace are shown, and a 
complete table of wavelengths at all the detected stationary points with respect to their field 
dependence is given, thereby providing a basis for a comparison with observed absorption spectra of 
magnetic white dwarfs.
\end{abstract}
\vspace{1cm}

\section{Introduction}

Since the astrophysical discovery of strong magnetic fields on the surfaces of white dwarfs ($\le 10^{5}$ 
Tesla) and neutron stars ($\approx 10^{8}$ Tesla) the interest in the behaviour and the properties of 
matter in strong magnetic fields has increased enormously. In the case of atoms in strong magnetic fields 
most of the literature is concerned with the hydrogen atom. The eigenvalues and the eigenfunctions of the 
hydrogen atom are, therefore, known very precisely and for many excited states 
\cite{friedrich,roesner,wintgen,ivanov88,kravchenko,ruder,schmelchschweiz}. For the hydrogen atom in a strong magnetic 
field it was possible to perform a comparison of the theoretical predictions resulting from {\it ab 
initio} computations with the data obtained from astronomical observation. The results provided an 
overwhelming evidence for the existence of hydrogen in the atmosphere of the corresponding astrophysical 
objects \cite{ruder}. However, there are several spectral features and structures which can not be 
explained by hydrogenic spectra like for example in the spectrum of the magnetic white dwarf GD229
\cite{green,schmidt,jssbecken}, which leads to the conjecture that there are further components in the 
atmospheres, i.e. atoms with more than one electron.

So far our knowledge about atoms with more than one electron in strong magnetic fields is very sparse. 
Most of the {\it ab initio} computations on more electron systems deal with two-electron systems, i.e. 
the helium atom, the hydrogen anion and the helium-like kations 
\cite{ruder,schlicht,proeschel,thurner,ivanov91}. For two-electron atoms the investigations cover much 
smaller parts of the spectrum than it is the case for the hydrogen atom, and their accuracy is 
considerably poorer. To perform a comparison with astrophysical observation, however, accurate transition 
energies for a large set of field strengths have to be available which requires even more accurate data
for the total energies. In particular in the relevant regime of 
intermediate field strengths, for which the diamagnetic energies and the Coulomb energies are of the 
same order of magnitude, there does not exist sufficiently accurate data for a large number of excited 
states which would allow this comparison. The reason is that for example the numerical basis set 
methods established so far have their starting point either in the low-field regime or in the 
high-field regime and are, therefore, especially adapted to their corresponding regimes but fail to be 
effective in the opposite regime, thereby leaving a gap in a certain regime where none of them provides 
a good convergence.

The scope of the present paper is to introduce one uniform basis set method which is capable of 
accurately describing two electron systems for arbitrary field strengths. We present numerical results 
for the energy levels of the ground state and a considerable number of excited states of the helium atom 
in a magnetic field ranging from $B=0$a.u. to $B=100$a.u. ($B=1a.u.$ corresponds to $2.35\cdot 10^5 
$Tesla).

In order to perform a detailed comparison of our numerical data with already existing ones, let us first 
provide an overview of the results present in the literature so far. Almost all of the results we mention 
here are variational, i.e. they provide upper bounds for the exact energy values. However, we have to 
distinguish between Hartree-Fock calculations and calculations which take into account the electronic 
correlation.

The first Hartree-Fock calculations on helium in a strong magnetic field have been performed in 1976 by 
Virtamo \cite{virtamo}. Virtamo has provided the global ground-state energies of helium in very strong 
fields ($B=80$a.u. up to $B=8 \times 10^4$a.u.). In this regime the lowest energy is achieved by aligning 
both spins antiparallel to the magnetic field which means that the global ground state in question is a 
spin triplet state. The same state is also investigated in the work of Pr\"oschel et al.\cite{proeschel}. 
They use a Slater-determinantal approach starting from Landau-levels in order to cover a slightly broader 
regime of field strengths as Virtamo ($B\sim 2.1$a.u. to $B\sim 2.1\times 10^4$a.u.). Hartree-Fock 
results for several other states besides the ground state of helium in the high-field regime are provided 
in the works of Ivanov \cite{ivanov91,ivanov94} where the singlet and triplet states are considered for 
positive total $z$-parity and magnetic quantum numbers $m=0,-1$ as well as for negative $z$-parity and 
$m=0$. Here the magnetic field ranges from $B=0$a.u. to $B=100$a.u. occupying 11 field strengths and thus
includes both the low-field and high-field regime as well as the intermediate regime. On an even finer
grid of field strengths (34 values between $B=8\times 10^{-4}$a.u. and $B=8\times 10^{3}$a.u.) the 
energies of the triplet ground state and several triplet excited states but no singlet states are given 
in the work of Thurner et al.\cite{thurner}. In the latter work there exists a gap in the list of 
eigenenergies for several states in the vicinity of $B\sim 1$a.u. because the ansatz of the applied 
approach changes in the intermediate regime from a spherical to a cylindrical symmetry and thus fails to 
provide accurate results in the intermediate regime. An important work is also the Hartree-Fock study of 
Jones, Ortiz and Ceperley \cite{ceperleyHF}. They present the HF-energies for the global low-field ground 
state for various field strengths between $B=8\times 10^{-4}$a.u. and $B=8$a.u. and thus also address the 
intermediate regime. They additionally provide energies for several excited states (but those results are 
crude approximations providing no upper bounds).

In comparison to the considerable number of Hartree-Fock investigations there are much less data 
available on fully correlated calculations of helium so far. Mueller, Rau and Spruch \cite{mueller} 
obtain variational upper-bound estimates of the binding energies of the triplet states with negative 
$z$-parity and the magnetic quantum numbers $m=0$ and $m=-1$ and of the singlet ground state which has 
positive $z$-parity and $m=0$. Their field strengths range from $B\sim 4.2$a.u. to 
$B\sim 2.1\times 10^4$a.u.. A similar high-field regime ($B=4.2$a.u.,$42$a.u.,$420$a.u.) is addressed by 
the work of Vincke and Baye \cite{vincke} which also presents variational estimates for binding energies. 
They consider the singlet and triplet states for positive $z$-parity and $m=0,-1,-2$. The first 
correlated calculations in the intermediate regime have been provided by Larsen \cite{larsen} who has 
given the energies of the singlet states with positive $z$-parity and $m=0,-1$ or with negative 
$z$-parity and $m=0$ as well as the triplet ground state for the four field strengths 
$B=0.2,0.5,1.0,2.0$a.u.. Park and Starace \cite{park} have computed upper and lower bounds for the 
singlet ground state in the low-field regime between $B=0$a.u. and $B=0.15$a.u.. Recently, Jones, Ortiz 
and Ceperley \cite{ceperleyRPQMC} have applied a released-phase quantum Monte Carlo method to the helium 
atom. Energies for spin triplet states with both positive and negative $z$-parity and $m=0,-1$ are given 
for field strengths from $B=0$a.u. up to $B=8.0$a.u.. Though in principle the results of those 
investigations are variational their numerical values must be handled with care when using them as upper 
bounds for the helium energies. The reason is that due to the statistical character of their method the 
results possess error bars which might trespass the exact values of the helium energies. Very recently 
the finite-element technique has as well been used to calculate energies for several singlet and triplet 
states of helium in a magnetic field\cite{braun98}. Also this approach does not provide a significant 
improvement.

In the present work we use a Gaussian basis set method which permits to perform {\it fully correlated} 
calculations on singlet and triplet states of helium with arbitrary spatial symmetry. This basis set 
method has been developped by Schmelcher and Cederbaum \cite{schmelcher} for molecules and has already 
successfully been applied to the hydrogen molecule ion \cite{kappes1} and to the neutral hydrogen 
molecule \cite{detmer,detmer98} in strong magnetic fields. We translate this basis set in order to be 
applicable to ab initio calculations of atoms in strong magnetic fields. We remark that using for example 
a Hylleraas basis set yields more accurate results in the field-free or weak-field case due to 
explicitely correlated orbitals. However, it is not efficient and accurate in the presence of a strong 
magnetic field due to the spherical symmetry of the exponentials \cite{scrinzi}. In contrast to this our 
basis set described in detail in sec. III below can be adapted to any field strength.

All the helium states considered are classified according to a maximal set of conserved quantities, which 
we choose to be the total spin $S^2$, the $z$-component $S_z$ of the total spin, the total spatial 
magnetic quantum number $M$ and the total spatial $z$-parity $\Pi_z$. For twenty field strengths from 
$B=0$a.u. up to $B=100$a.u. numerical helium energies for singlet and triplet states are given for $M=0$ 
and for $\Pi_z=0,1$. In each of these two subspaces we present the energies of the ground state and the 
first five excited states for singlet and four excited states for triplet spin symmetry. The accuracy of 
our method in the field-free case can be determined by comparison of our results with the very precise 
field-free values calculated by Braun et al.\cite{schweizer}, Accad et al.\cite{pekeris} or by Drake et 
al.\cite{drake}, and our relative deviation ranges from $10^{-6}$ to $10^{-4}$. We do not observe this 
accuracy to drop considerably with increasing magnetic field, which means that our basis set method 
produces accurate results in particular in the intermediate regime. Both with respect to the achieved 
relative accuracy as well as with respect to the number of excited states our investigation provides 
therefore valuable results. Data for $M\ne 0$ will be presented in a future work. The stationary 
components of the transitions obtained in the present investigation provide an important part of the data 
which have very recently been used to show the presence of helium in the atmosphere of the magnetic white 
dwarf GD229\cite{jssbecken}.

In detail we proceed as follows. In section II we discuss the influence of the finite nuclear mass, i.e. 
the scaling relations which connect the Hamiltonian for a finite nuclear mass to a Hamiltonian with 
infinite nuclear mass. In all of the remaining sections, we consider the nonrelativistic case with 
infinite nuclear mass and an electron spin $g$-factor equal 2. Section III starts with the 
nonrelativistic Hamiltonian of a helium atom with infinite nuclear mass in a magnetic field. The 
corresponding symmetries of this Hamiltonian are discussed and serve for constructing a suitable basis 
set. The matrix representation of the Hamiltonian with respect to this basis gives rise to an eigenvalue 
problem. The results of its diagonalization are discussed in section IV where we present total energies, 
ionization energies and transition energies as well as the wavelengths of their stationary components, 
all of them given under the assumption of infinite nuclear mass.

\section{Finite nuclear mass scaling relations and further corrections}

We use a nonrelativistic approach to the helium atom. This is justified because the relative changes in  
the helium energies due to relativistic effects are smaller than the relative accuracy of our 
nonrelativistic energies \cite{chen}. Additionally for simplicity we use an electron spin $g$-factor 
equal 2 throughout the paper. The following scaling laws as well as all of our numerical results can 
trivially be adapted to a $g$-factor of any desired accuracy by multiplying every occuring spin operator 
or eigenvalue with $\frac{g}{2}$.

Though our electronic structure calculations will be carried out with the assumption of an infinite 
nuclear mass, they can, by means of a scaling law, be translated into results with a finite nuclear mass. 
The latter are necessary in order to achieve the accuracy for a detailed comparison with astrophysical 
data. The idea to introduce such a scaling relation is not new but has already been applied to 
one-electron systems earlier \cite{pavlov}. In order to be demonstrate how this idea is applied on the 
helium atom, we will start with the full Hamiltonian describing a neutral system with two interacting 
electrons with masses $1$ and charge $-1$ in the Coulomb field of a nucleus with charge two and 
{\it finite} mass $M_0$ in a magnetic field $\mbf{B}$. Assuming a vanishing pseudomomentum of the center 
of mass (see refs. \cite{johnson,schmelchqc,schmelchlandau} for a pseudoseparation of the center of mass 
motion for neutral systems in a magnetic field) one obtains the following exact pseudoseparated 
Hamiltonian (which is established in the symmetric gauge $\mbf{A}(\mbf{r}) = \hf \mbf{B}\times\mbf{r}$)
\begin{eqnarray}
	H_e      &=&   \sum\limits_i^2 \left(    \frac{1}{2\mu}  \mbf{p}_i^2 
		   		               + \frac{1}{2\mu'}  \mbf{B}\cdot\mbf{l}_i
				               + \frac{1}{8\mu}  (\mbf{B}\times\mbf{r}_i)^2   
				               - \frac{2}{|\mbf{r}_i|} + \mbf{B}\cdot\mbf{s}_i 
				       \right)  
                     + \frac{1}{|\mbf{r}_2 - \mbf{r}_1|}                                     \nonumber \\
	         &&  + \frac{1}{2M_0}\sum\limits_{i\ne j} 
                        \left(   \mbf{p}_i\mbf{p}_j
			       - \mbf{p}_i(  \mbf{B}\times \mbf{r}_j)
			       + \frac{1}{4}  (\mbf{B}\times \mbf{r}_i)(\mbf{B}\times \mbf{r}_j)
                        \right) 
                                                                             \;\;\;, \label{ham_fin_mass}
\end{eqnarray}
where $\mu = \frac{M_0}{M_0+1}$ is the reduced mass and $\mu' = \frac{M_0}{M_0-1}$ (we use atomic units 
throughout the paper). We take into account the dominant part of the finite mass corrections by 
introducing these quantities $\mu,\mu'$ and neglect the mass polarization terms represented by the sum 
over $(i\ne j)$. Our basic idea is now find a scaling law joining the spectrum of the remaining part 
$H(M_0,B)$ in the first line of the Hamiltonian $H_e$ with the spectrum of the infinite-mass Hamiltonian 
$H(\infty,\bar{B})$ at a suitable different field strength $\bar{B}$. The neglect of the mass 
polarization terms of the sum over $(i\ne j)$ in eq.(\ref{ham_fin_mass}) is justified by the fact 
that they are expected to provide a smaller correction to the total energy than the corresponding 
diagonal mass corrections, i.e. the replacement of the masses by reduced ones.

We will show how it is possible to establish a relation between the operators $H(M_0,B)$ and 
$H(\infty,\bar{B})$ themselves which is even more fundamental than only a relation between the 
corresponding spectra. The relation in question cannot be a simple Unitarian because such a 
transformation would leave the spectrum invariant. Instead, we will show that there is an unique way how 
to represent $H(M_0,B)$ in the form $UH(M_0,B)U^{-1} = \alpha H(\infty,\bar{B}) + \beta$, where $\alpha$ 
is a number and $\beta$ is a suitable operator commuting with the Hamiltonian.

Before factoring out a suitable prefactor $\alpha$ the terms of $H(M_0,B)$ must be transformed in such a 
way that their mutual ratios possess the correct infinite-mass values, i.e. all ratios must neither 
contain $\mu$ nor $\mu'$. The misproportion between the kinetic energy and the Coulomb energy terms is 
the only one which cannot be absorbed in the parameter $B$ or be removed by separating a suitable 
additive operator $\beta$ commuting with the Hamiltonian. Thus we are obliged to introduce the canonical 
scale transformation $\mbf{r}\rightarrow \frac{1}{\mu}\mbf{r}$, $\mbf{p}\rightarrow \mu\mbf{p}$ of the 
coordinates themselves. We realize this transformation by the Unitarian 
$U=e^{-i\frac{\ln \mu}{2} (\mbf{x}\mbf{p}+ \mbf{p}\mbf{x})}$, yielding
\begin{eqnarray}
     UH(M_0,B)U^{-1}  &=&  \mu\left[\sum\limits_i^2 
			      \left(    \frac{1}{2}  \mbf{p}_i^2 
				      + \frac{1}{8}  (\smf{1}{\mu^2}\mbf{B}\times\mbf{r}_i)^2   
                                      - \frac{2}{|\mbf{r}_i|} 
                              \right) 
                         + \frac{1}{|\mbf{r}_2 - \mbf{r}_1|}\right]
		         + \sum\limits_i^2 
                              \left(    \frac{1}{2\mu'}  \mbf{B}\cdot\mbf{l}_i+ \mbf{B}\cdot\mbf{s}_i 
                              \right)
\end{eqnarray}
The operator $\mbf{s}_i$ of the spin degree of freedom is not affected by the above transformation, and 
$\mbf{l}_i=\mbf{r}_i\times\mbf{p}_i$ is also unchanged because the scaling factors of $\mbf{r}_i$ and 
$\mbf{p}_i$ cancel each other. The expression in square brackets is already an essential part of the 
desired infinite-mass Hamiltonian at the adjusted field strength $\bar{B} = \frac{1}{\mu^2}B$: only the 
Zeeman orbital and spin term at the field strength $\bar{B}$ are missing. These operators can be provided 
by hand, which must be repaired by subtracting the same operators at the end where they almost cancel 
exactly the original Zeeman and spin terms leaving only a contribution of order $\smf{B}{M_0}$:
\begin{eqnarray}
     UH(M_0,B)U^{-1}  &=&  \mu\left[\sum\limits_i^2 \left(  \frac{1}{2}  \mbf{p}_i^2 
							  + \frac{1}{2} \bar{\mbf{B}}\cdot\mbf{l}_i
				    			  + \frac{1}{8}  (\bar{\mbf{B}}\times\mbf{r}_i)^2   
							  - \frac{2}{|\mbf{r}_i|} 
				    			  + \bar{\mbf{B}}\cdot\mbf{s}_i 
						    \right) 
					+ \frac{1}{|\mbf{r}_2 - \mbf{r}_1|}
			      \right]                                                        \nonumber \\
		      &&    + \sum\limits_i^2 \left( - \mu\frac{1}{2} \bar{\mbf{B}}\cdot\mbf{l}_i 
                                                     - \mu \bar{\mbf{B}}\cdot\mbf{s}_i 
                                                     + \frac{1}{2\mu'}  \mbf{B}\cdot\mbf{l}_i
                                                     + \mbf{B}\cdot\mbf{s}_i 
                                              \right)
\end{eqnarray}
Therefore we obtain
\vspace{4mm}

\hspace{2.2cm}
\framebox{
\parbox{13cm}{
\begin{eqnarray}
     UH(M_0,B)U^{-1}  &=&  \mu \cdot H(\infty,B/\mu^2) 
                         - \frac{1}{M_0} \mbf{B}\cdot \sum\limits_i(\mbf{l}_i+\mbf{s}_i)    \label{scale}
\end{eqnarray}
}}
\vspace{2mm}

\noindent
In particular, the spectrum of $H(M_0,B)$ is identical with the spectrum of the unitarily equivalent 
operator represented by the right side of (\ref{scale}). The latter spectrum can be simply connected to 
the spectrum of $H(\infty,B/\mu^2)$ itself because the operators $\sum_i\mbf{B}\cdot \mbf{l}_i$ and 
$\sum_i\mbf{B}\cdot\mbf{s}_i$ commute with the Hamiltonian: we will use in the following section an 
angular momentum and spin basis, in the case of which the additional operator 
$-\frac{1}{M_0} \mbf{B}\cdot \sum\limits_i(\mbf{l}_i+\mbf{s}_i)$ gives rise to a trivial energy shift of 
order $\smf{B}{M_0}$.

\section{Symmetries, Hamiltonian and Basis sets}\label{sec_ham_bas}

\subsection{Symmetries and Hamiltonian}

In the following we assume the magnetic field to point in the $+z$-direction. Then the Hamiltonian 
$H(\infty,B)$ for infinite nuclear mass at given field strength $B$ (in the following we will omit the 
argument $'\infty'$) reads
\begin{eqnarray}
	H  &=&   \sum\limits_{i=1}^2 \left(   \hf\mbf{p}_i^2 +\hf B{{l}_z}_i 
				            + \smf{B^2}{8}(x_i^2+y_i^2) 
		                            - \smf{2}{|\mbf{r}_i|} 
                                            + B{{s}_z}_i 
				     \right)
		 \;\;+\;\;   \frac{1}{|\mbf{r}_2 - \mbf{r}_1|}                                \label{ham}
\end{eqnarray}
The sum contains the one-particle operators, i.e. the Coulomb potential energies $-\smf{2}{|\mbf{r}_i|}$ 
of the electrons in the field of the nucleus as well as their kinetic energies, already splitted into the 
part $\hf \mbf{p}_i^2$, the Zeeman term $\hf B{{l}_z}_i $ and the diamagnetic term 
$\smf{B^2}{8}(x_i^2+y_i^2)$, and their spin energies $B{{s}_z}_i$. The two-particle operator 
$\frac{1}{|\mbf{r}_2 - \mbf{r}_1|}$ represents the electron-electron repulsion energy.

There exist four independent commuting conserved quantities: the total spin $\mbf{S}^2$, the 
$z$-component $S_z$ of the total spin, the $z$- component $L_z$ of the total angular momentum and the 
total spatial $z$-parity $\Pi_z$. In the following calculations we consider separately each subspace of a 
specified symmetry, i.e. with given eigenvalues of $\mbf{S}^2$, $S_z$, $L_z$ and $\Pi_z$.

\subsection{The underlying one-particle basis set}\label{sec_onebas}

The key ingredient of our basis set method is the Gaussian one-particle basis set which is our starting 
point for the construction of spatial two-particle states. According to the azimuthal symmetry with 
respect to the magnetic field axis, cylindrical coordinates are suitable for representing the 
one-particle basis functions
\begin{eqnarray}
	\Phi_i(\rho,\varphi,z)  &=&     \rho^{{n_\rho}_i} z^{{n_z}_i} 
                                     e^{-\alpha_i\rho^2-\beta_i z^2} 
                                     e^{im_i\varphi}  \;\;\;\;i=1,...,n\;\;\;,             \label{onebas}
\end{eqnarray}
where $\alpha_i$ and $\beta_i$ are positive nonlinear variational parameters and the exponents 
${n_\rho}_i$ and ${n_z}_i$ obey the following restrictions:
\begin{eqnarray}
	{n_\rho}_i  &=& |m_i|     +  2k_i\;;\;\;\; k_i = 0,1,2,...\;\;with\;\;m_i = ...-2,-1,0,1,2,...
										         \label{gerade}\\ 
	{n_z}_i     &=& \;{\pi_z}_i\; +  2l_i\;;\;\;\;\; l_i = 0,1,2,...\;\;with\;\;{\pi_z}_i = 0,1 
                                                                                             \label{zpar}
\end{eqnarray}
The basis function $\Phi_i$ is an eigenfunction of the $z$-component of the angular momentum with 
eigenvalue $m_i$ and an eigenfunction of the $z$-parity with eigenvalue $(-1)^{{\pi_z}_i}$. The 
Gaussian-like expression  $\rho^{|m_i|}e^{-\alpha_i\rho^2}$ is identical with the $\rho$-dependence of 
the lowest Landau-state in the field $B$ if we choose $\alpha_i$ to be $\smf{B}{4}$ and thus represents 
an adjustment to the existence of the magnetic field (see e.g. eq. (8) in \cite{schmelchlandau}). The 
monomials $\rho^{2k_i}$ and $z^{{n_z}_i}$ are suitable for describing excitations. The flexibility of our 
basis set which permits us to choose suitably the values of the nonlinear parameters $\alpha_i$ and 
$\beta_i$ is one of its major advantages: For low field strengths an isotropic choice 
$\alpha_i = \beta_i$ will be reasonable where the $\alpha_i$, $\beta_i$ cover a regime which allows the 
$\Phi_i$ to optimally approximate Slater-type orbitals. At high field strengths, however, where the 
magnetic field destroys the spherical symmetry of the problem, an isotropic basis set method would be 
inefficient. Here we choose the distribution of the $\alpha_i$ to be peaked around $\smf{B}{4}$ whereas 
the $\beta_i$ are well-tempered in a large regime. 

The best choice of the $\alpha_i$ and $\beta_i$ has been computed by the requirement to solve optimally 
the one-particle problem of the $H$-atom or the He$^+$-ion in a magnetic field of given strength $B$. To 
achieve this, we applied the following optimization procedure. For any given one-particle subspace 
$(m,\pi_z)$ which will be involved in the later on two-particle configurations we have chosen a suitable 
number, typically 20, of functions of the type (\ref{onebas}) with the same one-particle quantum numbers 
$(m_i=m,{\pi_z}_i=\pi_z)$ but different starting values $\alpha_i$ and $\beta_i$. With respect to this 
basis we built up the overlap matrix and the matrix of the one-particle Hamiltonian, representing a 
generalized eigenvalue problem for the one-particle energies. Now, we determined systematically the 
values of the $\alpha_i$ and $\beta_i$ which minimized the energy eigenvalues of the one-particle ground 
state or a desired excited state. Our tool was a repeated reconstruction and diagonalization of the 
matrices directed by the pattern-search-method applied in the $\{\alpha_i,\beta_i\}$ space. We remark 
that due to the large number of parameters it is very time consuming to find the best selection of 
parameters $\{k_i,l_i\}$. Additionally the starting values for the nonlinear parameters $\alpha_i$ and 
$\beta_i$ have to be chosen carefully in order to find a ''good'' minimum close to the global one on the 
complicated hypersurface.

\subsection{The two-particle basis set}

Our basic idea for solving the time-independent Schr\"odinger equation for the Hamiltonian (\ref{ham}) is 
to construct a basis set of suitable two-particle states $\left.|q\right\rangle $ each of which is itself 
an eigenstate of the conserved quantities with eigenvalues $M,\Pi_z,S,S_z$, respectively. With respect to 
this not necessarily orthonormal basis we construct a matrix representation $H_{pq}$ of the Hamiltonian 
and define an overlap matrix $S_{pq}$ by 
\begin{eqnarray}
	H_{pq} &:=&  \left\langle p|H|q \right\rangle                                 \label{matrixelH}\\
	S_{pq} &:=&  \;\,\left\langle p\,|\,q \right\rangle                           \label{matrixelS} 
\end{eqnarray}
where the states $\left.|p\right\rangle$, $\left.|q\right\rangle$ lie in the {\it same} subspace of given 
eigenvalues of $M,\Pi_z,S,S_z$. By construction, the matrix $\underline{\underline{H}}$ is hermitian and 
the matrix $\underline{\underline{S}}$ is hermitian and positive definite. Additionally to these general 
properties, in the special case of the basis sets we use, all the matrix elements turn out to be real. By 
solving the finite-dimensional generalized real-symmetric eigenvalue problem
\begin{eqnarray}
	(\underline{\underline{H}} - E \underline{\underline{S}})\cdot \underline{c}  &=& 0     
											\label{seculareq}
\end{eqnarray}
we obtain eigenvectors $\underline{c}$ whose corresponding eigenvalues $E$ are strict upper bounds to the 
exact eigenvalues of the Hamiltionian (\ref{ham}) within the given subspace.

We choose our two-particle basis functions to be direct products of a pure spatial part 
$\left.|\psi_q\right\rangle$ with a pure spin part $\left.|\chi_q\right\rangle$:
\begin{eqnarray}
	\left.|q\right\rangle  &=& \left.|\psi_q\right\rangle \otimes \left.|\chi_q\right\rangle 
\end{eqnarray}
The spin part is assumed to be one of the usual orthonormal singlet or triplet spin eigenstates. This 
means that the overlap matrix is nontrivial purely due the spatial part,
\begin{eqnarray}
	S_{pq}   &=&    \left\langle \psi_p|\psi_q \right\rangle\;\;,
\end{eqnarray}
and each matrix element of the Hamiltonian decomposes in a spatial and a spin contribution:
\begin{eqnarray}
	H_{pq}   &=&            \left\langle \psi_p|H_{spat}|\psi_q \right\rangle  
                       + S_{pq} \left\langle \chi_p|H_{spin}|\chi_q \right\rangle           \label{decomp}
\end{eqnarray}
The spin contribution $\left\langle \chi_p|H_{spin}|\chi_q \right\rangle $ is trivial and equals 
$+B\cdot S_z$, according to (\ref{ham}), and therefore it vanishes for singlet states or is either 
$-B, 0$ or $+B$ for triplet states. Due to its prefactor $S_{pq}$ it can simply be absorbed as a shift in 
the energy eigenvalue $E$ in eq.(\ref{seculareq}). Since a spin singlet state is antisymmetric and a 
triplet state is symmetric with respect to particle exchange, the spatial part associated to a spin 
singlet state must be symmetric whereas the spatial part associated to a spin triplet state must be 
antisymmetric.

In contrast to the spin part the spatial part is by no means trivial and it is an art to find a suitable 
finite-dimensional basis set for accurately approximating the Hilbert space of bound states of 
$H_{spat}$. This is the reason why we spent so much effort on providing a powerful one-particle basis set 
(see sec. \ref{sec_onebas}). It serves now as a good starting point to compose the desired two-particle 
states. We choose a two-particle basis state to be
\begin{eqnarray}
	\left.|\psi_q\right\rangle &:=& b_i^{\dag} b_j^{\dag}\left. |0\right\rangle   
		\;\;\;\;i=1,...,n\;\;\;,\;\;\;\;j=i,...,n\;\;\;,                           \label{twobas}
\end{eqnarray}
where $ b_i^{\dag}$ is a creation operator of the $i$-th one-particle basis state 
$\left.|i\right\rangle =  b_i^{\dag}\left. |0\right\rangle$ whose position representation is given by 
(\ref{onebas}). This means that we will treat the helium atom by a {\bf full Configuration Interaction 
(full CI)} approach within the one-particle basis set of states $\left.|i\right\rangle$. Depending on 
whether the spin part $\left.|\chi_q\right\rangle $  is a singlet or triplet state, the operators 
$ b_i^{\dag}$ must be bosonic or fermionic, respectively.

Now, in order to establish a basis set of two-particle states spanning the subspace with the given pure 
values of the total magnetic quantum number $M$ and the total $z$-parity $\Pi_z$, we must select all the 
combinations $i,j$ with
\begin{eqnarray}
	             m_i+m_j &=& M               \\
mod({\pi_z}_i + {\pi_z}_j,2) &=& \Pi_z  \;\;\;,
\end{eqnarray}
yielding a dimension $N$ of the constructed two-particle basis set which is in general smaller than 
$\smf{n(n+1)}{2}$. We remark that for triplet spin symmetry those states in (\ref{twobas}) with $i=j$ 
fail to exist which means that the dimension $N$ for triplet subspaces is in general smaller than for 
singlet subspaces.

\subsection{Matrix elements}

In order to calculate the matrix elements of the Hamiltonian (\ref{ham}), we must rewrite its spatial 
part in second quantization, $\hat{H}_{spat}  =  \hat{H}_{I}  +  \hat{H}_{\roemII}$, where $\hat{H}_{I}$ 
and $ \hat{H}_{\roemII}$ denote the second-quantized counterparts of the familiar one- and two-particle
operators whose position representations read
\begin{eqnarray}
	H_{I}(\mbf{p},\mbf{r})            &=&      \hf \mbf{p}^2 + \hf \mbf{B}\cdot\mbf{l} 
                                                 + \smf{1}{8}(x^2+y^2)  - \smf{2}{|\mbf{r}|}           \\
	H_{\roemII}(\mbf{r}_1,\mbf{r}_2)  &=&      \smf{1}{|\mbf{r}_2 - \mbf{r}_1|} 
\end{eqnarray}
The next step now is to calculate the spatial matrix elements according to (\ref{decomp}). With 
    $\left.|\psi_q\right\rangle := b_i^{\dag} b_j^{\dag}\left. |0\right\rangle$ 
and $\left.|\psi_p\right\rangle := b_k^{\dag} b_l^{\dag}\left. |0\right\rangle$ a straightforward 
calculation leads to
\begin{eqnarray}
        \left\langle \psi_p | \psi_q \right\rangle \;\;\;            
&=&     \;\;\;\;\;\;  \left\langle i | k \right\rangle \;\;\left\langle j | l \right\rangle \;\;\;\;\,
     \pm    \;\;\;\;  \left\langle i | l \right\rangle\;\; \left\langle j | k \right\rangle    
											\label{matrel1}\\
	\left\langle \psi_p |\hat{H}_{I}| \psi_q \right\rangle       
&=&           \;\;\;  \left\langle i | H_{I} | k \right\rangle   \; \left\langle j | l \right\rangle  
     \;\;\; \pm \;\,  \left\langle i | H_{I} | l \right\rangle \;   \left\langle j | k \right\rangle  
											     \nonumber \\
&&	    +         \left\langle j | H_{I} | l \right\rangle \;   \left\langle i | k \right\rangle 
     \,\;\; \pm  \;   \left\langle j | H_{I} | k \right\rangle \;   \left\langle i | l \right\rangle   \\ 
        \left\langle \psi_p |\hat{H}_{\roemII}| \psi_q \right\rangle 
&=&     \;\;\;\;\;\;  \left\langle ij | H_{\roemII} | kl \right\rangle \;\;\;\;
	\pm \;\;\;\;  \left\langle ij | H_{\roemII} | lk \right\rangle  \;\;\;\;,         \label{matrel3}
\end{eqnarray}
where $\left.|ij\right\rangle := \left.|i\right\rangle \otimes \left.|j\right\rangle$ and where the sign 
'$\pm $' stands for '$+$' in the singlet case and for '$-$' in the triplet case.

All the $\frac{n(n+1)}{2}$ different one-particle matrix elements 
$\left\langle i | H_{I} | k \right\rangle$ are relatively easily evaluated (see Appendix 
\ref{appendixB}). Each matrix elements $\left\langle i | H_{I} | k \right\rangle$ equals a prefactor 
depending on the parameters of $\left.| i \right\rangle$ and $\left.| k \right\rangle$ times the 
one-particle overlap $\left\langle i | k \right\rangle$, reflecting the fact that the one-particle 
operators do not only conserve the total magnetic quantum number $M$ and the total $z$-parity $\Pi_z$ but 
also the magnetic quantum numbers and $z$-parities of the individual one-particle states. We remark that 
the mentioned prefactors in the matrix elements of the Zeeman term do not even depend on any basis state,
rendering them purely proportional to the overlap, 
$\left\langle i | \hf \mbf{B}\cdot\mbf{l} | k \right\rangle = \hf m_k B \left\langle i | k \right\rangle$. 
Thus the Zeeman term just gives rise to a shift in the energies.

The only coupling between two-particle states composed of one-particle states with different combinations 
of magnetic quantum numbers or $z$-parities arises due to the two-particle operator of the 
electron-electron interaction. In contrast to the one-particle matrix elements the evaluation of the 
matrix elements $\left\langle ij | H_{\roemII} | kl \right\rangle$ is by no means trivial (see Appendix 
\ref{appendixC}). An additional problem is the large number of different two-particle matrix elements 
which is of the order $\frac{N(N+1)}{2}$ rather than $\frac{n(n+1)}{2}$. This means that a sophisticated 
and detailed analysis of the analytical representation of the two-particle matrix elements by means of 
series of different representations of hypergeometric functions has been necessary in order to achieve an 
effective numerical implemetation and an acceptable CPU time for performing a calculation for one given 
field strength. Details on the representation of the quantities 
$\left\langle ij | H_{\roemII} | kl \right\rangle$ are given in Appendix \ref{appendixC}.

\section{Results and discussion}

\subsection{Spectroscopic notation and properties}

Before presenting the numerical results of our calculations we shall explain our spectroscopic notation 
in the presence of the field as well as its correspondence to the field-free notation. According to the 
four conserved quantities $M$, $\Pi_z$, $\mbf{S}^2$ and $S_z$ we denote a state by 
$\nu^{2S+1}_{S_z}M^{(-1)^{\Pi_z}}$ where $(2S+1)$ is the spin multiplicity and $\nu =1,2,3,...$ is the 
degree of excitation within a given subspace. If obvious, we will omit the index $S_z$ in the following. 
In the present paper we will investigate and show results for the subspaces ${}^10^+$, ${}^30^+$, 
${}^10^-$, ${}^30^-$.

For vanishing field, there exists a one-to-one correspondence between our field notation and the common 
field-free notation $n^{2S+1}_{S_z}L_M$ (see Table 1). It is given by the fact that for bound helium 
eigenstates of $\mbf{L}^2$ and $L_z$ we have $(-1)^{\Pi_z} = (-1)^{L+M}$. The latter relation is not in 
general true for multi-electron systems because for a configuration containing two angular momenta 
$l_1,l_2$ we have $(-1)^{\Pi_z} = (-1)^{l_1+l_2+M}$. For helium and $B=0$, however, all doubly excited 
states lie in the continuum\cite{westerveld} which means that all the helium states below the 
one-particle ionization threshold $T(B=0) = -2.0 a.u.$ must be one-particle excitations with $l_1=0$ and 
$l_2=L$.

We emphasize that the mentioned one-to-one correspondence for $B=0$ does not contradict the fact that the 
$\mbf{L}^2$-symmetry is higher than the ${\Pi_z}$-symmetry: At a finite field strength the former is 
broken whereas the latter is still valid. The field free notation becomes meaningless for finite field
and could only be maintained as a labelling device for the energy eigenstates because a given 
$\nu^{2S+1}_{S_z}M^{(-1)^{\Pi_z}}$ state at a finite field develops in a unique way from the equally 
labelled $\nu^{2S+1}_{S_z}M^{(-1)^{\Pi_z}}$ state at $B=0$ which, in turn, is identical with one unique 
$n^{2S+1}_{S_z}L_M$ state. We remark that energy curves within a subspace of given symmetries 
${}^{2S+1}_{S_z}M^{(-1)^{\Pi_z}}$ are not expected to cross \cite{neumann}, whereas crossings between 
curves of different subspaces are allowed.

In Table 1 we also provide the energies of the field free states which have been very precisely 
calculated by Accad et al. \cite{pekeris} or by Drake et al. \cite{drake}. We want to keep track of the 
energetical order of the states for several reasons. The first reason is that we need the energies for 
associating the energy quantum numbers $\nu$ to the states. It would not be sufficient to use the same 
counting $n$ as in the field free case because there exist different states with the same $n$ but 
different $L$ which possess the same ${\Pi_z}$-symmetry, as is evident already for the states $3^10^+$ 
and $4^10^+$. 

The second reason is to point out the approximate degeneracy of the field free states with the same 
energy quantum number $n$ but different $L$. Whereas the electron-electron interaction is only able to 
slightly perturb the otherwise exact degeneracy of those states, we will see that the magnetic field will 
completely remove this degeneracy. This effect can even be observed for the behaviour of the 
corresponding states in the H atom and thus is primarily an effect of the magnetic field alone rather 
than of the two-particle character of the He atom. It occurs in addition to the well-known removal of the 
degeneracy of states with the same quantum number $L$ but different $M$ whose energies split for finite 
field in $(2L+1)$ different energy values.

The properties of the field free states of helium summarized in Table 1 serve as a good starting point 
for presenting our data for finite fields.

\subsection{Aspects for the selection of basis functions}

In order to obtain accurate results by our basis set method two major difficulties have been overcome by 
an optimal selection of basis functions. The first difficulty is the limited number $n$ of one-particle 
functions from eq.(\ref{onebas}) which can be used to describe the exact wave function. The second 
difficulty, which is not completely independent from the previous one, is to describe electronic 
correlation by using a basis composed of one-particle states.

One manifestation of correlation is the fact that different electrons avoid occupying the same region in 
space which we expect to be the consequence of the electron-electron repulsion as well in a magnetic field. 
In particular, we decribe this by two electrons tending to occupy two regions in opposite directions from 
the nucleus which corresponds to the situation $\varphi_2-\varphi_1 = \pi$. The Coulomb interaction 
$\frac{1}{|\mbf{r}_1-\mbf{r}_2|}$ breaks the independent conservation of the $z$-components 
${l_z}_1,{l_z}_2$ of the two angular momenta, only leaving the sum $L_z = {{l_{z}}_1} + {l_z}_2$ as 
conserved quantity. A Fourier representation of the angular part of any fully correlated two-particle 
wave function $\Psi(\mbf{r}_1,\mbf{r}_2)$ is given by 
$\Psi(\mbf{\varphi}_1,\mbf{\varphi}_2) = \sum_{m_1,m_2}A_{m_1m_2}e^{i(m_1\varphi_1 + m_2\varphi_2)}$. In 
order to obtain an eigenfunction of $L_z$ we must demand the constraint $m_1+m_2 = M$, yielding
$\Psi(\mbf{\varphi}_1,\mbf{\varphi}_2) = e^{iM\varphi_1}\sum_{m}A_{M-m,m}e^{im(\varphi_2-\varphi_1)}$. 
This expression contains angular correlation as can already be seen explicitely by considering the lowest 
cosine term $B_m\cos m(\varphi_2-\varphi_1)$ contained in the series above: assuming $B_m$ to be 
negative, $\Psi(\mbf{\varphi}_1,\mbf{\varphi}_2)$ is largest for $\varphi_2-\varphi_1 = \pi$ and lowest 
for $\varphi_2-\varphi_1 = 0$. Therefore, it appears fruitful to choose one-particle wave functions with 
opposite magnetic quantum numbers $\pm 1$ and $\pm 2$ or even higher angular momenta combining to a total 
magnetic quantum number $M=0$ in order to describe well the angular correlation. Correlation can also 
been described by wave functions possessing a node at the nucleus position and thus allowing the two 
electrons to be located in opposite directions with respect to the nucleus.

The maximum dimension $N$ of the Hamiltonian matrix is limited by CPU and storage resources. Through a 
very efficient implementation (see appendices \ref{appendixA},\ref{appendixB},\ref{appendixC}) of the 
matrix elements and by optimized storage usage we were able to push the number of two-particle basis 
functions to $N\approx 4300$ which represents the limit with respect to linear dependencies of similar 
basis functions resulting in instabilities of the numerical diagonalisation of the generalized eigenvalue 
problem (\ref{seculareq}).

Let us discuss the strategy for the selection of basis functions for the example of the subspace $0^+$. 
We first focus on the basis functions optimized for a nuclear charge $Z=1$, i.e. the hydrogen atom. We 
selected 31 functions with the symmetry $0^+$, among which 13 have a $\rho$-exponent equal 2 instead of 
$0$ in order to describe correlation with the aid of their nodes. For further improving the description 
of correlation we used also each 13 functions with $m^{\pi_z}=\pm1^{+}$ and $m^{\pi_z}=\pm2^{+}$ as well 
as each 13 functions with $m^{\pi_z}=0^{-}$ and $m^{\pi_z}=\pm1^{-}$. The first excited state is already
rather well described by this basis set but is still improved by adding basis functions optimized for 
excitations of the H atom in a magnetic field. Whereas for the first excited state correlation effects 
are still important the higher excitations are, like in the field free case, more and more dominated by 
one-electron excitations for which automatically the two electrons are spatially separated. Therefore,
higher excitations were throughout described by adding functions with values $k_i=1$ or $l_i=1$ in 
eqs.(\ref{gerade}) and (\ref{zpar}) which are subsequently optimized to describe the corresponding higher 
excitations within the $0^+$ subspace of hydrogen. 

In order to describe higher excitations of the He atom the optimization for $Z=1$ is sufficient because 
the nucleus with $Z=2$ is screened by the inner electron. For the ground state and the first two excited 
states, however, we have observed important contributions also from functions optimized for $Z=2$. Here 
we used a similar selection scheme as for the functions optimized for $Z=1$, but only half the number of 
functions.

Altogether we arrive at a number of $n=244$ different one-particle states, from which $N_1=4378$ 
two-particle states for singlet and $N_3=4288$ two-particle states for triplet spin symmetry can be 
composed according to eq.(\ref{twobas}).

In a similar manner, we built up the basis for the $0^-$ subspace by involving functions optimized for 
$Z=1$ as well as for $Z=2$. We observe the effect of the latter optimized functions to be less 
significant in the $0^-$ case than in the $0^+$ case. This appears natural since in any two-particle 
state with $0^-$ symmetry at least one electron is in an excited one-particle state and is only 
attracted by a screened nucleus with effective charge closer to $Z=1$ than to $Z=2$. Due to this reason 
we used altogether $n=195$ one-particle states which is slightly less than in the $0^+$ case. According 
to eq.(\ref{twobas}) we obtain $N_1 = N_3 = 3600$ two-particle states for the singlet and triplet 
subspace, respectively. The numbers $N_1$ and $N_3$ do not differ due to the fact that the identical 
spatial one-particle quantum numbers which are always forbidden for the triplet two-particle states do 
not occur for the $0^-$ singlet subspace either because two different one-particle $z$-parities are 
required to form an odd total $z$-parity.

In the following subsection we discuss the results of our He calculations. Their accuracy is estimated 
by comparison with the field free data. If available, we provide also a comparison of our data for finite 
field strengths with the literature. 
\vspace{2mm}

\subsection{Energies for finite field strengths}

\subsubsection{Results for $M=0$ and even $z$-parity}

{\bf a) Results for the singlet states $\nu^10^+$}

For the singlet subspace $\nu^10^+$ we present the ground state and the first four excitations, i.e. 
$1\le \nu \le 5$. In the low-field and part of the intermediate regime the $1^10^+$ state is the global 
ground state. The numerical results for the total energy of the $1^10^+$ state as a function of the 
magnetic field are shown in Table 2. We observe that apart from $B=0$ and $B=0.08$ our results for the 
energies are throughout lower and thereby better than the best ones given in the literature. We remark 
that the $1^10^+$ state is the state which causes the most difficulties for an accurate description by 
our method. The first reason is that it is the only state for which both electrons considerably occupy 
the one-particle ground state which forces the two electrons to be close to each other in a narrow domain 
of space and which thus gives rise to a relatively strong contribution of electronic correlation. The 
second reason is that the mentioned one-particle ground state, which would be a Slater-type orbital for 
$B=0$, possesses a cusp at the origin which is difficult to obtain accurately by a superposition of 
Gaussians. This second effect, however, can be expected to become less important with increasing field 
strength since the cusp in the direction perpendicular to the field axis is smoothened out by the 
increasing dominance of the magnetic field.

The overall increase of the total energies (see Table 2) with increasing field strength has its origin in 
the strongly increasing kinetic energy in the presence of the external field. The total binding energy of 
the two electrons can be obtained from the total energies $E(B)$ in Table 2 by subtracting the minimal 
energy $B$ of two free electrons in the field, yielding $E(B) - B$ (here and in the following all 
ionizations are considered with fixed quantum numbers).

However, for an analysis of the electronic structure the one-electron ionization energies for the process 
He $\rightarrow$ He$^+ + e^-$ are much more sensitive. The corresponding one-particle ionization 
threshold $T(B)$ is provided in the fourth column of Table 2. This threshold $T$ can easily be obtained 
as the sum of the lowest Landau-energy $\frac{B}{2}$ of the ionized electron and the total one-particle 
energy $E^{(1)}_{tot}(Z=2)$ of the other electron in the Coulomb field of the nucleus with charge $Z=2$ 
in the presence of the magnetic field. This quantity $E^{(1)}_{tot}(Z=2)$, in turn, can be received from 
the highly accurate values for the one-particle binding energy $E^{(1)}_{bind}(Z=1)$ computed by 
Kravchenko et al. \cite{kravchenko}. First we extract the total energies for $Z=1$ from 
$(-E^{(1)}_{bind}(Z=1)):= E^{(1)}_{tot}(Z=1)- \frac{B}{2}$, and then we use the nuclear charge scaling 
relation $E(Z,B) = Z^2 E(Z=1,B/Z^2)$ in ref.\cite{ruder}, yielding
\begin{eqnarray}
	T(B)  &=&  B - 4E^{(1)}_{bind}(Z=1,B/4) 
\end{eqnarray}
This threshold $T$ lies essentially closer to the values $E(1^10^+)$ than the higher threshold $B$ for 
two-particle ionization. We consider it in fact to be the optimal reference point for the total energies, 
and as expected the total energies of the excitations $\nu^10^+$ which are given in Table 3 approach 
closer and closer to the value $T$ with increasing excitation. In agreement with the mentioned 
considerations about the special difficulties in achieving accurate results for the ground state we 
obtain much preciser values for the excitations within the same basis set. 

The one-particle ionization energies for all the states $\nu^10^+$, $1\le \nu \le 5$, i.e. the values of 
$E(B)-T(B)$, are given in Fig.1 as a function of the magnetic field. A logarithmic scale on the energy 
axis is necessary for covering the three orders of magnitude for the different states. We observe that 
though the ground state $1^10^+$ becomes monotonically stronger bound with increasing field strength 
this is not in general the case for the excitations. Below $B{ \sim}0.005 a.u.$ none of the energies 
differs considerably from its field free value. Between $B{ \sim}0.005 a.u.$ and $B{ \sim}0.1 a.u.$ a 
rearrangement takes place which is caused by the increasing dominance of the magnetic forces over the 
Coulomb forces. The first effect to be observed is that the $4^10^+$ state leaves the energetical 
vicinity of the $3^10^+$ state. According to Table 1, for $B=0$ these states would coincide with the 
states $3^1S_0$ and $3^1D_0$, respectively, and thus both correspond to the same energy quantum number 
$n=3$. The degeneracy of these two states is only slighty disturbed in the field-free case whereas a 
finite field removes this approximate degeneracy completely above $B{ \sim}0.01 a.u.$. 

At $B{ \sim}0.08 a.u.$ the states $4^10^+$ and $5^10^+$ experience an avoided crossing. It can be 
recognized well even though the strong curvature of the graphs would make a finer grid of calculated 
energy values on the field axis desirable. The present grid is, nevertheless fine enough to observe that 
the calculated points are well aligned, confirming the good convergence of our calculations.
\vspace{3mm}

\noindent
{\bf b) Results for the triplet states $\nu^30^+$}

For the triplet subspace $\nu^30^+$ we present the ground state and the first three excitations, i.e. 
$1\le \nu \le 4$. The numerical results together with the available data from the literature are given 
in Table 4. Much more data are available in the literature than for the singlet states. These numbers 
convincingly demonstrate that our approach to the solution of the two-electron problem in a strong 
magnetic field is superior to any other method existing in the literature.

Among the three related triplet states with $S_z = 0,\pm1$ the one with $S_z=-1$ possesses the lowest 
energy due to the spin shift $BS_z$. Therefore, we have given in Table 4 those lowest energies. Although
for $B=0$ the singlet state $1^10^+$ is the global ground state and the triplet state $1^3_00^+$ (i.e. 
$S_z=0$) stays above it for any field strength, the spin shift $(-B)$ causes the related triplet state 
$1^3_{-1}0^+$($S_z=-1$) to become the ground state within the $M=0$ subspace above $B\sim 1.112a.u.$. For 
the triplet ground state it is within our approach much easier to obtain accurate results than for the 
singlet ground state. This can be understood in the following picture: in the triplet ground state only 
one of the electrons occupies the one-particle ground state with the cusp whereas the other one occupies 
already predominantly an excited one-particle state and thus gives rise to a lower correlation 
contribution than in the singlet ground state.

In analogy to the case of the singlet states we provide in Fig.1 the dependence of the one-particle 
ionization energies on the field strength for the triplet states. The reference threshold $T(B)$ has 
the same value as for the singlet case because, following our general definition of the threshold, we 
keep the spins fixed. 

We observe that the ionization energy of any state $\nu^30^+$ in Fig.1 behaves similarly to the 
ionization energy of its singlet counterpart $(\nu+1)^10^+$. This relationship has its origin in the fact 
that two states differing only with respect to their spin symmetry possess according to the field free 
notation in Table 1 similar contributions to their energies. The only difference with respect to the 
matrix elements of eqs.(\ref{matrel1})-(\ref{matrel3}) for two such states is the sign of the exchange 
terms. Those might be small if the particles are sufficiently spatially separated as it is the case for 
all states apart from the $1^10^+$ state. This statement is confirmed by the fact that the 
singlet-triplet splitting is particularly small between the $4^10^+$ state and the $3^30^+$ state. These 
states belong to the field free states $3^1D_0$ and $3^3D_0$ (see Table 1), respectively, and the 
one-particle orbitals with $d$-symmetry involved here are even more spatially separated from the 
participating $s$-orbitals than the $p$-orbitals playing a role in other states with larger 
singlet-triplet splitting. Of course this argumentation breaks down when the magnetic field destroys the 
spherical symmetry sufficiently and causes a mixture of total angular momenta. Consequently, the 
splitting between the $4^10^+$ state and the $3^30^+$ state grows considerably at $B\sim 0.02$.

\subsubsection{Results for $M=0$ and odd $z$-parity}

{\bf a) Results for the singlet states $\nu^10^-$}

For the singlet subspace $\nu^10^-$ we present the ground state and the first four excitations, i.e. 
$1\le \nu \le 5$. For $B\le 0.08$, even the energies for the state $6^10^-$ show an excellent convergence. 
The corresponding data are presented in Table 5. For a graphical representation we choose the ionization 
energies which are calculated from the total energies with respect to the same threshold $T(B)$ as in the 
case of the $0^+$ subspace. The reason is that in a one-electron ionization process of helium constrained 
to keep $0^-$ symmetry the ionized electron can adopt the negative $z$-parity, allowing the remaining 
electron to occupy the one-particle ground state which possesses $0^+$ symmetry. The ionized electron, in 
turn, has the same Landau energy $\frac{B}{2}$ as in the $0^+$ case because the $z$-parity does not have 
any influence on the Landau energy which is only assigned to the transversal degrees of freedom.

The curves for the ionization energies are shown in Fig.2. We observe that for very low fields the
$6^10^-$ state and the $5^10^-$ state are approximately degenerate which is in agreement with the fact 
that for two values of $\nu,\mu$ within the same bracket of the sequence 
$(1),(2),(3,4),(5,6),(7,8,9),(10,11,12)...$ the two states $\nu^10^-$ and $\mu^10^-$ correspond to the 
same quantum number $L$ according to Table 1. As in the $0^+$ subspace, this approximate degeneracy is 
removed for fields above $B\sim 0.02$. The region between $B\sim 0.02$ and $B\sim 0.1$ exhibits some 
avoided crossings.
\vspace{3mm}

\noindent
{\bf b) Results for the triplet states $\nu^30^-$}

In contrast to the singlet case there exist much more data in the literature for the triplet state. In 
Table 6 we have listed them together with our results for the states $\nu^3_{-1}0^-$ (i.e. $S_z = -1$), 
$1\le \nu \le 5$. Again our results are better than any references for finite field strengths.

The ionization energies for the $\nu^3_{-1}0^-$ states are also shown in Fig.2. As in the singlet case, 
we achieved a good convergence even for the fifth excited state $6^30^-$ below $B= 0.08$. The 
singlet-triplet splitting between the states $\nu^10^-$ and $\nu^30^-$ behaves similar as a function of 
the field like the corresponding splitting in the ${}^{1/3}0^+$ subspaces.

\subsection{Transitions}

In order to interpret experimental spectra from magnetic white dwarfs like GD229, it is necessary to 
determine transition energies from our total energies. The selection rules for electric dipole 
transitions are $\Delta S = 0$, $\Delta S_z = 0$ for the spin degrees of freedom and 
$\Delta M = 0, \;\Delta \Pi_z = \pm 1$ or $\Delta M = \pm 1, \;\Delta \Pi_z = 0$ for the spatial degrees 
of freedom. With the data presented we are able to treat the $\Delta M = 0$ transitions. We obtain a 
spectrum of 30 transition energy curves for singlet transitions and 24 ones for triplet transitions which 
we show in Fig.3a and 3b, respectively. In both logarithmic representations, we observe singularities 
belonging to zeros in the transition energies which arise due to level crossings of the initial state 
and the final state.

Since the magnetic field of a white dwarf is not constant but varies by a factor of two for a dipole 
geometry, the spectrum of wavelengths is in general smeared out. Transitions whose wavelengths are 
stationary with respect to the magnetic field, however, reflect themselves by characteristic absorption 
edges in the observable spectrum. 

Due to this important role of the stationary lines we have summarized all transitions showing stationary 
points in Tables 7 and 8. The position and the wavelength of each stationary point have been determined 
by interpolation using the relatively crude grid of our calculations. This allows us to perform a 
comparison of the stationary components with the observed spectrum of the magnetic white dwarf GD229
\cite{jssbecken}. Indeed, the accurate data on many excited states presented here are part of an analysis 
accomplished very recently \cite{jssbecken} which clearly shows that there is strong evidence for the 
existence of He in the atmosphere of GD229. This might also be the case for other magnetic white dwarfs 
and therefore the present data will serve astrophysicists for further comparison with observational data.

\section{Conclusions and Outlook}

We have investigated the electronic structure of the helium atom in a magnetic field by a fully 
correlated approach. We assumed the nucleus to possess infinite mass but provided scaling laws how to 
connect our fixed-nucleus results to the data which should be expected for the true finite nuclear mass. 
One of the goals of our work, the identification of the features in the spectrum of the white dwarf GD229 
with electronic transitions in atomic helium, has already successfully been demonstrated\cite{jssbecken}. 
This has been possible due to the high accuracy of our calculations on many excited states considering 
the electronic structure for magnetic fields for $0\le B \le 100a.u.$, i.e. 
$0\le B \le 2.3505 \cdot 10^7$ Tesla.

The starting point of our {\it ab initio} treatment of the helium atom in a magnetic field was its full 
fixed-nucleus Hamiltonian which possesses four conserved quantities: the total spin $\mbf{S}^2$ and its 
$z$-component $S_z$, the $z$-component $L_z$ of the electronic angular momentum and the electronic 
$z$-parity $\Pi_z$. Due to the magnetic field the spherical rotational invariance is broken resulting in 
the fact that the total electronic angular momentum $\mbf{L}^2$ fails to be a conserved quantity for 
nonvanishing field $B\ne 0$. 
Our full configuration interaction approach has been able to overcome the difficulty that the symmetry of 
the system changes from purely spherical for $B=0$ to mainly cylindrical for high fields. To this purpose 
we used a one-particle basis set of anisotropic Gaussians furnished with monomials in the coordinates 
$\rho$ or $z$ transversal or longitudinal to the field axis, respectively. The degree of anisotropy has 
been obtained as a result from the direct optimization of the nonlinear parameters for the transversal or 
longitudinal Gaussian by the requirement to solve optimally the one-particle problem of the H atom or the 
He${}^+$ ion in a magnetic field of given strength. The one-particle basis functions were composed to 
two-particle states of pure total symmetry in the sense of the four conserved quantities mentioned above. 
In the present paper, we have provided results for the singlet and triplet states in the subspaces with 
vanishing spatial magnetic quantum number $M=0$ and positive or negative $z$-parity.  

In each of these subspaces we have built up a matrix for the Hamiltonian and for the overlap between the 
nonorthogonal two-particle states which provided a variational estimation for the energy eigenvalues 
after the diagonalization of a generalized eigenvalue problem. By very elaborate techniques for the 
evaluation of the matrix elements we were able to treat matrix dimensions of about 4000 within less than 
one day CPU time on a moderate Silicon Graphics workstation. The relative accuracy of the results for 
the energies ranged between $10^{-4}$ for the singlet ground states and $10^{-5}$ for the triplet 
states or general excited states for $2<\nu<5$. To achieve this, a careful selection of the basis 
functions was necessary, i.e. combinations of one-particle magnetic quantum numbers $m=\pm1,\pm2,...$ 
were important to describe correlation whereas one-particle functions describing high excitations of the 
one-particle systems H or He${}^+$ provided the major contribution also for high excitations of the 
two-particle system He.

To enable a comparison of our calculated transitions with the observed spectra of magnetic white dwarfs, 
we have determined all the stationary points associated to the transitions between states with $M=0$ and 
different $z$-parities. It turned out that the stationary points in the regime $B\sim 0.1$ and 
$B\sim 0.3$ can be identified with lines in the spectrum of the white dwarf GD229 which is even further 
confirmed by transitions involving $M=-1$ and positive $z$-parity\cite{jssbecken,0minus1becken}. This 
provides strong evidence for the existence of helium in the atmosphere of this white dwarf. The present 
data will additionally be helpful in order to see whether there is evidence for helium also in other 
magnetic cosmic objects.

In order to complete the treatment of the electronic structure of helium in a magnetic field, we will in 
the near future calculate data for subspaces with  higher magnetic quantum numbers for both odd and even 
$z$-parity. Furthermore, a detailed investigation of the oscillator strengths as a function of the 
magnetic field is necessary in order to estimate properly the intensities of the transitions. 
\vspace{5mm}

{\it Acknowledgements.} The Deutsche Studienstiftung (W.B.) and the Deutsche Forschungsgemeinschaft 
(W.B.) are gratefully acknowledged for financial support. P.S. acknowledges many helpful discussions 
during the CECAM workshop on 'Atoms in strong magnetic fields'. F.K.D. acknowledges financial support by 
the European Union. It is a pleasure for us to thank U. Kappes for many fruitful discussions as well as 
for his input with respect to the computational aspects of the present work.

\begin{appendix}

\section{The overlap matrix elements}\label{appendixA}

In the following we only give the results for the analytic representation of the matrix elements. The 
evaluation of the overlap integral is very simple because it factors into an integral over the 
transversal coordinate $\rho$ and an integral over the longitudinal coordinate $z$. Both integrals can be 
reduced to a Gaussian type $\int\limits_0^\infty t^n e^{-\gamma t^2},\, n\ge 0$, yielding for the 
dependence of the one-particle overlap $\left\langle i | k \right\rangle$ on the parameters of the states 
$| i \rangle$ and $| k \rangle$:
\begin{eqnarray}
     \left\langle i | k \right\rangle &=&   \delta_{m_i m_k} \delta_{{\pi_z}_i{\pi_z}_k} \,\cdot\,
					    \pi^{\frac{3}{2}} 
					    \frac{  ({n_z}_{ik}-1)!! \; (\frac{{n_{\rho}}_{ik}}{2})!}
                                                 {  2^{\frac{{n_z}_{ik}}{2}} 
                                                    \alpha_{ik}^{\frac{{n_{\rho}}_{ik}}{2}+1}
                                                    \beta_{ik}^{\frac{{n_z}_{ik}+1}{2}}}      \label{ovl}
\end{eqnarray}
where we define $(-1)!! := 1$ and $\gamma_{ik} := \gamma_i + \gamma_k$. The Kronecker Delta symbols 
reflect the orthogonality of two one-particle states with different magnetic quantum numbers or different 
$z$-parities. We observe that the overlap is real and all the parameters of the wave functions 
$| i \rangle$ and $| k \rangle$ enter symmetrically in eq.(\ref{ovl}).

\section{The one-particle matrix elements}\label{appendixB}

\subsection{The matrix elements of the kinetic energy}

The matrix elements of the operator $(\mbf{p} + \frac{1}{2}\mbf{B}\times\mbf{r})^2$ are evaluated by 
using cylindric coordinates in which we consider
\begin{eqnarray}
     (\mbf{p} + \frac{1}{2}\mbf{B}\times\mbf{r})^2 
&=& - \frac{1}{2}\left(   \left(    \frac{\partial^2}{\partial \rho^2}
                                  + \frac{1}{\rho} \frac{\partial}{\partial \rho}
                          \right) 
                        + \frac{1}{\rho^2}\frac{\partial^2}{\partial \varphi^2}
        		+ \frac{\partial^2}{\partial z^2}
			+ iB \frac{\partial}{\partial \varphi}
			- \frac{1}{4}B^2\rho^2 
     		 \right)                                                                     \nonumber \\
&=:& \left(T_{\rho}+T_{\varphi}+T_z  + T_{Zeeman} +  T_{dia} \right)\;,
\end{eqnarray}
where $T_{\rho},T_{\varphi},T_z$ represent the Laplacian, $T_{Zeeman}$ is the Zeeman term and $T_{dia}$ 
the diamagnetic term. The evaluation of any of the matrix elements 
$\left\langle i | T_i | k \right\rangle$ is relatively simple since the derivatives generate prefactors 
but leave the types of the integrals unchanged in comparison to the overlap integral. A consequence is 
that the results are found to be proportional to the overlap $\left\langle i | k \right\rangle$ between 
the {\it same} two states. In detail we have
\begin{eqnarray}
      \left\langle i | T_{\rho}| k \right\rangle  
&=&   \left\langle i | k \right\rangle    \cdot 
      \left\{ \begin{array}{ll} \frac{  -   {n_{\rho}}_k^2\alpha_i^2 
                                        + 2({n_{\rho}}_i {n_{\rho}}_k 
                                        +   {n_{\rho}}_{ik})\alpha_i\alpha_k 
                                        -   {n_{\rho}}_i^2\alpha_k^2}
	                             {  {n_{\rho}}_{ik} \alpha_{ik} }    &;\;\; {n_{\rho}}_{ik} \ne 0  \\
                                \frac{2\alpha_i\alpha_k}{\alpha_{ik}}    &;\;\; {n_{\rho}}_{ik} = 0  
              \end{array}
      \right\}                                                                                         \\
      \left\langle i | T_{\varphi} | k \right\rangle   
&=&   \left\langle i | k \right\rangle    \cdot 
      \left\{ \begin{array}{cl} \frac{\alpha_{ik}}{{n_{\rho}}_{ik}}m_k^2 &;\;\; {n_{\rho}}_{ik} \ne 0  \\
                        0                                                &;\;\; {n_{\rho}}_{ik} = 0  
              \end{array}
      \right\}                                                                                         \\
      \left\langle i | T_z | k \right\rangle   
&=&   \left\langle i | k \right\rangle    \cdot 
      \left\{ \begin{array}{ll} \frac{      {n_z}_k(1-{n_z}_k)\beta_i^2 
                                        + (2{n_z}_i{n_z}_k 
                                        +   {n_z}_{ik} - 1)\beta_i\beta_k 
                                        +   {n_z}_i(1-{n_z}_i)\beta_k^2}
				     {     ({n_z}_{ik} - 1)\beta_{ik}}   &;\;\; {n_{z}}_{ik} \ne 1     \\
              \end{array}
      \right\}                                                                                         \\
      \left\langle i | T_{Zeeman} | k \right\rangle   
&=&   \left\langle i | k \right\rangle    \cdot \frac{1}{2}m_k B                                       \\
      \left\langle i | T_{dia} | k \right\rangle   
&=&   \left\langle i | k \right\rangle    \cdot 
                                \frac{{n_{\rho}}_{ik} + 2}{2\alpha_{ik}} \cdot \frac{1}{8}B^2
\end{eqnarray}
Again we observe that each matrix element is real and symmetric with respect to an interchange of the 
states $| i \rangle$ and $| k \rangle$. Due to the selection rules in the prefactor 
$\left\langle i | k \right\rangle$ the occurence of the single quantity $m_k$ does not destroy this 
symmetry. Physically, the proportionality to $\left\langle i | k \right\rangle$ means that the operator 
of the kinetic energy does not couple one-particle states involving different magnetic quantum number or 
$z$-parity.

\subsection{The matrix elements of the electronic Coulomb interaction with the nucleus}

The evaluation of the matrix elements of the one-particle operator $V_I=\frac{1}{\mbf{r}}$ is much more 
complicated than the other one-particle matrix elements discussed above. The reason is that the Coulomb 
potential possesses a spherical symmetry rather than the cylindrical symmetry of the basis functions. We 
have overcome this difficulty by applying a Singer transformation\cite{singer}, leaving for the spacial 
integrations the convenient Gaussian types. The remaining integration due to the Singer transformation, 
however, is not as simple but it can be solved by involving the Gaussian hypergeometric function 
${}_2F_1$, yielding
\begin{eqnarray}
      \left\langle i | V_I | k \right\rangle 
&=& - \left\langle i | k \right\rangle \cdot\; 
         \beta_{ik}^{\frac{1}{2}}\;
         \frac{\Gamma\left(\frac{{n_{\rho}}_{ik}+{n_{z}}_{ik}}{2} + 1\right)}
       	      {\Gamma\left(\frac{{n_{\rho}}_{ik}+{n_{z}}_{ik}}{2} + \frac{3}{2}\right)}
	 \;\;{}_2F_1\left(  \smf{{n_{\rho}}_{ik}}{2} + 1\,,
                          \;\smf{1}{2}\,,
                          \;\smf{{n_{\rho}}_{ik}+{n_{z}}_{ik}}{2}+\smf{3}{2}\,;
                          \;1-\smf{\beta_{ik}}{\alpha_{ik}}\right)                     \label{matrelcoul}
\end{eqnarray}
It is important to remark that very elaborate techniques are necessary to evaluate the function ${}_2F_1$ 
for the various occuring arguments with a high accuracy and with an acceptable efficiency. The standard 
expansion of ${}_2F_1(a,b,c;z)$ in a power series of $z$ is by no means sufficient because a singularity 
of ${}_2F_1(a,b,c;z)$ for $z=1$ constrains the convergence domain of such a series to $|z|<1$. The 
consequence is that for $\frac{\beta_{ik}}{\alpha_{ik}}\ll1$ the convergence of the standard series would 
be arbitrarily slow. Moreover, for $\frac{\beta_{ik}}{\alpha_{ik}}>2$ the application of the standard 
series is completely useless. Using basis sets whose parameters $\alpha_i,\beta_i$ usually cover the 
range from $10^{-4}$ up to $10^5$, we have thus been obliged to use various formulas for suitable 
analytic continuations of ${}_2F_1(a,b,c;z)$ to the domain $|z|\ge 1$ \cite{abramowitz}. By these 
techniques we achieved an accuracy of $10^{-13}$ for the one-particle interaction integrals with no loss 
of efficiency compared with the simple integrals of the kinetic energy.

We remark that also the matrix elements of the electron-nucleus interaction are proportional to the 
overlap $\left\langle i | k \right\rangle$. This is in agreement with the fact that the spherically 
symmetric Coulomb potential $V_I=\frac{1}{\mbf{r}}$ neither breaks the azimuthal symmetry nor disturbs 
the behaviour of the one-particle basis functions under the reflection $z \rightarrow -z$.

\section{The two-particle matrix elements}\label{appendixC}

In contrast to the one-particle matrix elements treated above we cannot provide a single formula for the 
efficient and accurate calculation of the matrix elements 
$\left\langle ij | H_{\roemII} | kl \right\rangle$ of the two-particle interaction. The reason is the 
large number of different cases due to the various constellations of the individual one-particle 
functions which have to be treated differently in order to ensure a fast and highly accurate evaluation. 
We therefore provide in the following only an outline of the procedure and indicate the necessary 
techniques. Details can be obtained from the authors upon request.

First we apply a Singer transformation\cite{singer} in order to remove the Coulomb singularity in the 
integrand of the matrix elements, thereby introducing the new variable $u$ according to 
$\frac{1}{|\mbf{r}_1-\mbf{r}_2|}=\frac{2}{\sqrt{\pi}}\int_0^{\infty}du\;e^{-u^2(\mbf{r}_1-\mbf{r}_2)^2}$.
Although the underlying symmetry of the basis set is cylindrical, it is advantageous to carry out the 
spatial integrations in Cartesian coordinates. The $z_1z_2$-integral can always be factored out, and the 
coupling between the two particles can be removed by the subsitution 
$z_1\rightarrow z_1 - \frac{u^2}{\beta_{ik}+u^2}$. The resulting integral can be decomposed into a sum 
of $({n_z}_{ik} + 1)$ integrals each factoring in two pure $z_1$ and $z_2$ integrals of the Gaussian 
type $\int_{-\infty}^{\infty}z^{n_z}e^{-\beta(u)z^2}$ which can easily be evaluated.

The transversal part of the electron-electron integral is much more complicated than the $z$-integration. 
First each term $\rho_1^{|m_i|+|m_k|+2k_{ik}}e^{-i(m_i-m_k)\varphi_1}$ gives rise to factors
$(k_{ik}+1)\cdot (|m_i|+1) \cdot (|m_k|+1)$ 
according to its decomposition to 
$(x_1^2+y_1^2)^{k_{ik}}(x_1-i\;\mbox{sgn}(m_i)y_1)^{|m_i|}(x_1+i\;\mbox{sgn}(m_k)y_1)^{|m_k|}$ in 
Cartesian coordinates (again the same number of expressions arises for particle 2). Next, the particle 
decoupling transformation analoguous to the one mentioned above multiplies the number of integrals by an 
additional large factor $n_d$.

At this stage, all the Gaussian integrations in $x_1,x_2,y_1,y_2,z_1$ and $z_2$ can be carried out, 
resulting in functions of the remaining variable $u$ of type 
$u^2(1+a_1u^2)^{r_1}(1+a_2u^2)^{r_2}(1+a_3u^2)^{r_3}(1+a_4u^2)^{r_4}$, where the $r_i$ are positive or 
negative integers or half-integers. The coordinates can always be chosen such that one $r_i$ is a 
positive integer. Multiplying out the factor $(1+a_iu^2)^{r_i}$ enables us to reduce the remaining 
$u$-integral to an expression involving the Appell hypergeometric function $F_1(a,b,b',c;t_1,t_2)$. 

For each of the $\frac{N(N+1)}{2}$ matrix elements $\left\langle ij | H_{\roemII} | kl \right\rangle$, we 
thus have to evaluate 
$n_d \cdot \{(k_{ik}+1)\cdot (|m_i|+1)\cdot (|m_k|+1)\}^2 \cdot ({n_z}_{ik} + 1) \cdot (r_i+1)$ 
times the function $F_1$ with in general various different arguments $a,b,b',c$. Only the two arguments 
$t_1,t_2$ are universal for a fixed matrix element: $t_1 = \frac{\beta_{ik}}{\beta_{ik}+\beta_{jl}}$, and 
$t_2 = 1 - 
\frac{(\alpha_{ik}+\alpha_{jl})\beta_{ik}\beta_{jl}}{(\beta_{ik}+\beta_{jl})\alpha_{ik}\alpha_{jl}}$. For 
$|t_1|\ll 1$ and $|t_2|\ll 1$ we implemented the usual standard series expression of 
$F_1(a,b,b';c;t_1,t_2)=\sum_{m=0}^{\infty}\frac{(a,m)(b,m)}{(c,m)(1,m)}\;_2F_1(a+m,b';c+m;t_2)\;t_1^m$, 
where $_2F_1(a,b;c;z) = \sum\limits_{n=0}^{\infty}\;\frac{(a,n)(b,n)}{(c,n)(1,n)}\;z^n$ is the Gaussian 
hypergeometric function. However, the fact that for many basis functions either the argument $|t_1|$ or 
$|t_2|$ or even both happen to lie close to $1$ or above makes this standard representation useless since 
its convergence domains are the unit circles with respect to $t_1$ and $t_2$\cite{appell}. Therefore, it 
was necessary to use a large number of analytic continuation formulas for $F_1$ each of which is valid 
for one specified class of arguments $a,b,b',c$. Even more, we have systematically compared the CPU times 
for altogether more than 50 ways how to evaluate $F_1$ and selected the fastest one for each class of 
arguments, involving e.g. all of the continuation formulas for the Gaussian hypergeometric function 
${}_2F_1(a,b,c;z)$ given in eqs.(15.3.3-15.3.12) in ref.\cite{abramowitz}. We have additionally derived 
new formulas for ${}_2F_1(a,b,c;z)$ adapted for parameter constellations not covered in 
ref.\cite{abramowitz}, and we investigated the different possibilities how to reduce $F_1$ to $_2F_1$ or 
one of its derivatives.

We point out that without such a systematic analysis of the convergence properties of series 
representations for $F_1$ the present work would not have been possible: The reduction of CPU time for 
the same accuracy ($10^{-12}$) of our representation of the series compared to the most primitive 
selection of continuation formulas is about $10^3$.

\end{appendix}

\newpage

\end{document}


\begin{center}
{\LARGE\bf Tables}
\end{center}
\vspace{1cm}

\begin{table}
\begin{minipage}{16.9cm}
\caption{ Correspondence table between the field notation $\nu^{2S+1}M^{(-1)^{\Pi_z}}$ and the familiar 
field free notation $n^{2S+1}L_M$ for the lowest 55 singlet states and the lowest 54 triplet states of 
He. We have also provided the precise total zero-field energy of each state given in the literature. }
\begin{minipage}{7.0cm}
\vspace{1mm}

\begin{tabular}{|l|   l  ||           r          l         |}\hline\hline
\multicolumn{2}{|c||}{\rule[-2mm]{-2mm}{6mm} field free  }&\multicolumn{2}{|c|}{\rule[-2mm]{-2mm}{6mm}  field} \\ \hline
\multicolumn{1}{|c|} {\rule[-2mm]{-2mm}{6mm}   Energy    }&\multicolumn{1}{|c||}{\rule[-3mm]{0mm}{5mm} $n^{2S+1}L_M$} &\multicolumn{2}{|c|}{\rule[-3mm]{0mm}{5mm} $\nu^{2S+1}M^{(-1)^{\Pi_z}}$}\\ \hline
 $-2.903724377^a$ & $1^1S_0$       & $1^10^+$         &                      \\ \hline\hline
 $-2.145974046^a$ & $2^1S_0$       & $2^10^+$         &                      \\ \hline
$-2.1238430864^b$ & $2^1P_0      $ & $1^10^-$         &                      \\ 
                  & $2^1P_{\pm 1}$ & $1^1(\pm 1)^+$   &                      \\ \hline\hline
 $-2.061271988^a$ & $3^1S_0$       & $3^10^+$         &                      \\ \hline
$-2.0556207328^b$ & $3^1D_0      $ & $4^10^+$         &                      \\ 
                  & $3^1D_{\pm 1}$ & $1^1(\pm 1)^-$   &                      \\ 
                  & $3^1D_{\pm 2}$ & $1^1(\pm2)^+ $   &                      \\ \hline
$-2.0551463620^b$ & $3^1P_0      $ & $2^10^-$         &                      \\ 
                  & $3^1P_{\pm 1}$ & $2^1(\pm 1)^+$   &                      \\ \hline\hline
 $-2.033586717^a$ & $4^1S_0$       & $5^10^+$         &                      \\ \hline
$-2.0312798461^b$ & $4^1D_0      $ & $6^10^+$         &                      \\ 
                  & $4^1D_{\pm 1}$ & $2^1(\pm 1)^-$   &                      \\ 
                  & $4^1D_{\pm 2}$ & $2^1(\pm2)^+ $   &                      \\ \hline
$-2.0312551443^b$ & $4^1F_0      $ & $3^10^-$         &                      \\ 
                  & $4^1F_{\pm 1}$ & $3^1(\pm 1)^+$   &                      \\ 
                  & $4^1F_{\pm 2}$ & $1^1(\pm2)^- $   &                      \\ 
                  & $4^1F_{\pm 3}$ & $1^1(\pm3)^+ $   &                      \\ \hline
$-2.0310696504^b$ & $4^1P_0      $ & $4^10^-$         &                      \\ 
                  & $4^1P_{\pm 1}$ & $4^1(\pm 1)^+$   &                      \\ \hline\hline
 $-2.0211137^c$   & $5^1S_0$       & $7^10^+$         &                      \\ \hline
$-2.0200158361^b$ & $5^1D_0      $ & $8^10^+$         &                      \\ 
                  & $5^1D_{\pm 1}$ & $3^1(\pm 1)^-$   &                      \\ 
                  & $5^1D_{\pm 2}$ & $3^1(\pm2)^+ $   &                      \\ \hline
$-2.0200029371^b$ & $5^1F_0      $ & $5^10^-$         &                      \\ 
                  & $5^1F_{\pm 1}$ & $5^1(\pm 1)^+$   &                      \\ 
                  & $5^1F_{\pm 2}$ & $2^1(\pm2)^- $   &                      \\ 
                  & $5^1F_{\pm 3}$ & $2^1(\pm3)^+ $   &                      \\ \hline
$-2.0200007108^b$ & $5^1G_0      $ & $9^10^+$         &                      \\ 
                  & $5^1G_{\pm 1}$ & $4^1(\pm 1)^-$   &                      \\ 
                  & $5^1G_{\pm 2}$ & $4^1(\pm2)^+ $   &                      \\ 
                  & $5^1G_{\pm 3}$ & $1^1(\pm3)^- $   &                      \\ 
                  & $5^1G_{\pm 4}$ & $1^1(\pm4)^+ $   &                      \\ \hline
$-2.0199059899^b$ & $5^1P_0      $ & $6^10^-$         &                      \\ 
                  & $5^1P_{\pm 1}$ & $\hspace{5mm}6^1(\pm 1)^+$   &          \\ \hline
\end{tabular}
\end{minipage}
\hspace{15mm}
\begin{minipage}{7.0cm}
\vspace{1mm}

\begin{tabular}{|l|   l  ||           r          l         |}\hline\hline
\multicolumn{2}{|c||}{\rule[-2mm]{-2mm}{6mm} field free  }&\multicolumn{2}{|c|}{\rule[-2mm]{-2mm}{6mm}  field} \\ \hline
\multicolumn{1}{|c|}{\rule[-2mm]{-2mm}{6mm}   Energy }&\multicolumn{1}{|c||}{\rule[-3mm]{0mm}{5mm} $n^{2S+1}L_M$} &\multicolumn{2}{|c|}{\rule[-3mm]{0mm}{5mm} $\nu^{2S+1}M^{(-1)^{\Pi_z}}$}\\ \hline
\multicolumn{1}{|c|}{ --- }&\multicolumn{1}{|c||}{ --- }&\multicolumn{2}{|c|}{ --- }       \\ \hline\hline
 $-2.175229378^a$ & $2^3S_0$       & $1^30^+$         &                      \\ \hline
 $-2.133164191^b$ & $2^3P_0      $ & $1^30^-$         &                      \\ 
                  & $2^3P_{\pm 1}$ & $1^3(\pm 1)^+$   &                      \\ \hline\hline
 $-2.068689067^a$ & $3^3S_0$       & $2^30^+$         &                      \\ \hline
$-2.058081084^b$  & $3^3P_0      $ & $2^30^-$         &                      \\ 
                  & $3^3P_{\pm 1}$ & $2^3(\pm 1)^+$   &                      \\ \hline
$-2.055636309^b$  & $3^3D_0      $ & $3^30^+$         &                      \\ 
                  & $3^3D_{\pm 1}$ & $1^3(\pm 1)^-$   &                      \\ 
                  & $3^3D_{\pm 2}$ & $1^3(\pm2)^+ $   &                      \\ \hline\hline
 $-2.036512083^a$ & $4^3S_0$       & $4^30^+$         &                      \\ \hline
$-2.032324354^b$  & $4^3P_0      $ & $3^30^-$         &                      \\ 
                  & $4^3P_{\pm 1}$ & $3^3(\pm 1)^+$   &                      \\ \hline
$-2.031288847^b$  & $4^3D_0      $ & $5^30^+$         &                      \\ 
                  & $4^3D_{\pm 1}$ & $2^3(\pm 1)^-$   &                      \\ 
                  & $4^3D_{\pm 2}$ & $2^3(\pm2)^+ $   &                      \\ \hline
$-2.031255168^b$  & $4^3F_0      $ & $4^30^-$         &                      \\ 
                  & $4^3F_{\pm 1}$ & $4^3(\pm 1)^+$   &                      \\ 
                  & $4^3F_{\pm 2}$ & $1^3(\pm2)^- $   &                      \\ 
                  & $4^3F_{\pm 3}$ & $1^3(\pm3)^+ $   &                      \\ \hline\hline
 $-2.02261624^c$  & $5^3S_0$       & $6^30^+$         &                      \\ \hline
$-2.020551187^b$  & $5^3P_0      $ & $5^30^-$         &                      \\ 
                  & $5^3P_{\pm 1}$ & $\hspace{5mm}5^3(\pm 1)^+$   &          \\ \hline
$-2.020021027^b$  & $5^3D_0      $ & $7^30^+$         &                      \\ 
                  & $5^3D_{\pm 1}$ & $3^3(\pm 1)^-$   &                      \\ 
                  & $5^3D_{\pm 2}$ & $3^3(\pm2)^+ $   &                      \\ \hline
$-2.020002957^b$  & $5^3F_0      $ & $6^30^-$         &                      \\ 
                  & $5^3F_{\pm 1}$ & $6^3(\pm 1)^+$   &                      \\ 
                  & $5^3F_{\pm 2}$ & $2^3(\pm2)^- $   &                      \\ 
                  & $5^3F_{\pm 3}$ & $2^3(\pm3)^+ $   &                      \\ \hline
$-2.020000711^b$  & $5^3G_0      $ & $8^30^+$         &                      \\ 
                  & $5^3G_{\pm 1}$ & $4^3(\pm 1)^-$   &                      \\ 
                  & $5^3G_{\pm 2}$ & $4^3(\pm2)^+ $   &                      \\ 
                  & $5^3G_{\pm 3}$ & $1^3(\pm3)^- $   &                      \\ 
                  & $5^3G_{\pm 4}$ & $1^3(\pm4)^+ $   &                      \\ \hline
\end{tabular}

\end{minipage}
\end{minipage}
\vspace{1mm}

$^a$ Baker et al. (1990) \cite{baker90}

$^b$ Drake et al. (1992) \cite{drake}

$^c$ Accad et al. (1971) \cite{pekeris}

\end{table}

\begin{table}
\begin{minipage}{9.6cm}
\caption{ Total energies $E$ of the singlet ground state $1^10^+$, one-electron ionization threshold $T$ 
and, if available, best energy values given in the literature, as a function of the magnetic field 
strength $B$}
\vspace{1mm}

\begin{tabular}{|l|l|l|l|l|}\hline\hline
$B$     &  $ E(1^10^+)$  &\multicolumn{1}{|c|}{literature \rule[-3mm]{0mm}{8mm}}    &  $T$                \\ \hline
        &                &                    &                       \\
0.0000  &  $-2.903351$   &$-2.903724377^a$    & $-2.000      $        \\
0.0008  &  $-2.903346$   &                    & $-1.999599960$        \\    
0.004   &  $-2.903342$   &                    & $-1.997999000$        \\ 
0.008   &  $-2.903340$   &                    & $-1.995995995$        \\ 
0.020   &  $-2.903270$   &                    & $-1.989975001$        \\    
0.040   &  $-2.903036$   &                    & $-1.979900008$        \\    
0.080   &  $-2.902083$   &$-2.90195^d$        & $-1.959600176$        \\
0.160   &  $-2.898290$   &                    & $-1.918402804$        \\
0.240   &  $-2.892035$   &                    & $-1.876414090$        \\  
0.400   &  $-2.872501$   &$-2.8714^d$         & $-1.790105922$        \\
0.500   &  $-2.855859$   &$-2.85385^e$        & $-1.734628064$        \\   
0.800   &  $-2.787556$   &$-2.77585^d$        & $-1.561526260$        \\
1.000   &  $-2.729508$   &$-2.7272^e$         & $-1.440989741$        \\
1.600   &  $-2.507952$   &                    & $-1.058421519$        \\
2.000   &  $-2.329780$   &$-2.3266^e$         & $-0.788842154$        \\ 
5.000   &  $-0.574877$   &                    & $ 1.456132354$        \\
10.000  &  $ 3.064582$   &                    & $ 5.609851957$        \\
20.000  &  $11.267051$   &                    & $14.47840453 $        \\ 
50.000  &  $38.076320$   &                    & $42.45369755 $        \\ 
100.000 &  $84.918313$   &                    & $90.43945348 $        \\   \hline
\end{tabular}
\end{minipage}
\vspace{1mm}

$^a$ Baker et al. (1990) \cite{baker90}

$^d$ Thurner et al. (1982) \cite{schlicht}

$^e$ Larsen (1979) \cite{larsen}
\end{table}

\begin{table}
\begin{minipage}{15.8cm}
\caption{ Total energies $E$ of the excited singlet states $\nu^10^+$ for $2\le \nu\le 5$ as a function 
of the magnetic field strength $B$ as well as field free reference values}
\vspace{1mm}

\begin{tabular}{|l|l|l|l|l|}\hline\hline
$B$            &       $ E(2^10^+)$       &       $ E(3^10^+)$          &         $ E(4^10^+)$        &       $ E(5^10^+)$                \\ \hline
               &                          &                             &                             &                                   \\                         
0.0000         &$-2.145912     $          &$-2.061255     $             &$-2.055613           $       &$-2.033579        $                \\
0.0000 (lit.)  &$-2.145974046^a$          &$-2.061271988^a$             &$-2.055620732852245^b$       &$-2.033586717026^a$                \\
0.0008         &$-2.145911     $          &$-2.061246     $             &$-2.055607           $       &$-2.033531        $                \\    
0.004          &$-2.145869     $          &$-2.061028     $             &$-2.055489           $       &$-2.032867        $                \\ 
0.008          &$-2.145743     $          &$-2.060374     $             &$-2.055122           $       &$-2.031451        $                \\ 
0.020          &$-2.144852     $          &$-2.056838     $             &$-2.052030           $       &$-2.026808        $                \\    
0.040          &$-2.141787     $          &$-2.050850     $             &$-2.038691           $       &$-2.018836        $                \\    
0.080          &$-2.130812     $          &$-2.036626     $             &$-2.004123           $       &$-1.995610        $                \\
0.160          &$-2.096743     $          &$-1.999728     $             &$-1.962448           $       &$-1.945128        $                \\
0.240          &$-2.054961     $          &$-1.956971     $             &$-1.919910           $       &$-1.902875        $                \\  
0.400          &$-1.965307     $          &$-1.865411     $             &$-1.830581           $       &$-1.814978        $                \\
0.500          &$-1.908671     $          &$-1.807640     $             &$-1.773700           $       &$-1.758191        $                \\   
0.800          &$-1.736422     $          &$-1.632339     $             &$-1.599198           $       &$-1.584566        $                \\
1.000          &$-1.617870     $          &$-1.511771     $             &$-1.478551           $       &$-1.464143        $                \\
1.600          &$-1.241729     $          &$-1.130213     $             &$-1.096276           $       &$-1.081100        $                \\
2.000          &$-0.975861     $          &$-0.861382     $             &$-0.826959           $       &$-0.811912        $                \\ 
5.000          &$ 1.252363     $          &$ 1.379993     $             &$ 1.416706           $       &$ 1.432270        $                \\
10.000         &$ 5.393200     $          &$ 5.530945     $             &$ 5.569418           $       &$ 5.585430        $                \\
20.000         &$14.248991     $          &$14.396817     $             &$14.437004           $       &$14.453455        $                \\ 
50.000         &$42.207510     $          &$42.368689     $             &$42.411075           $       &$42.428159        $                \\ 
100.000        &$90.180690     $          &$90.351969     $             &$90.395965           $       &$90.414700        $                \\   \hline
\end{tabular}
\end{minipage}
\vspace{1mm}

$^a$ Baker et al. (1989) \cite{baker89}

$^b$ Drake et al. (1992) \cite{drake}
\end{table}

\begin{sidewaystable}
{\small
\begin{minipage}{18.2cm}
\caption{ Total energies $E$ of the triplet states $\nu^3_{-1}0^+$ ($S_z=-1$) for $1\le \nu\le 4$ as a 
function of the magnetic field strength $B$. We have also provided the results given in the literature so 
far.}
\vspace{1mm}

\begin{tabular}{|l||l      |l                ||l               |l                    ||l               |l                     ||l             |l                     | }\hline\hline
\rule[-3mm]{0mm}{8mm}
        &\multicolumn{2}{|c||}{ $E(1^30^+)$} &  \multicolumn{2}{|c||}{ $E(2^30^+)$}   &  \multicolumn{2}{|c||}{ $E(3^30^+)$}   &  \multicolumn{2}{|c|}{ $E(4^30^+)$}         \\ \hline
\multicolumn{1}{|c||}{ $B$ }
        &  this work       &     literature  &  this work       &     literature      &  this work      &     lit.             &  this work      &     literature            \\ \hline
        &                  &                 &                  &                     &                 &                      &                 &                           \\
0.0000  &   $-2.175220$    &   $-2.175229$   &  $-2.068687$     &  $-2.068689$        &   $-2.055630$   & $$                   &   $-2.036511$   & $-2.036512$               \\
        &                  &$\;\;\;\;\;378^a$&                  &$\;\;\;\;\;067472^a$ &\multicolumn{2}{|c||}{$-2.055636309453^b$} &              & $\;\;\;\;\;083098^a$      \\
0.0008  &   $-2.176019$    &   $-2.175050^f$ &  $-2.069480$     &  $-2.069278^i$      &   $-2.056424$   &                      &   $-2.037278$   &                           \\
0.004   &   $-2.179190$    &   $-2.178220^f$ &  $-2.072505$     &  $-2.072302^i$      &   $-2.059507$   &                      &   $-2.039893$   &                           \\
0.008   &   $-2.183098$    &   $-2.1830(5)^g$&  $-2.075969$     &  $-2.0760(1)^g$     &   $-2.063153$   &                      &   $-2.042333$   & $-2.0413(4)^g$            \\
0.020   &   $-2.194461$    &                 &  $-2.084626$     &                     &   $-2.072654$   &                      &   $-2.048567$   &                           \\
0.040   &   $-2.212236$    &   $-2.211222^f$ &  $-2.096580$     &  $-2.092686^i$      &   $-2.083143$   &                      &   $-2.060290$   &                           \\
0.080   &   $-2.243958$    &   $-2.2438(3)^g$&  $-2.121107$     &  $-2.1209(1)^g$     &   $-2.087409$   &                      &   $-2.082249$   & $-2.0687(9)^g$            \\
0.160   &   $-2.296318$    &                 &  $-2.166290$     &                     &   $-2.123762$   &                      &   $-2.105446$   &                           \\
0.240   &   $-2.339560$    &                 &  $-2.206689$     &  $-2.1927^f$        &   $-2.162609$   &                      &   $-2.143909$   &                           \\
0.400   &   $-2.412723$    &   $-2.4112^f$   &  $-2.279523$     &  $-2.2615^f$        &   $-2.235815$   &                      &   $-2.217383$   &                           \\
0.500   &   $-2.454347$    &   $-2.4528^h$   &  $-2.321985$     &                     &   $-2.279382$   &                      &   $-2.261303$   &                           \\ 
0.800   &   $-2.573615$    &   $-2.5737(3)^g$&  $-2.443352$     &  $-2.4395(9)^g$     &   $-2.403631$   &                      &   $-2.386891$   & $-2.3497(21)^g$           \\
1.000   &   $-2.650655$    &   $-2.6492^h$   &  $-2.520902$     &                     &   $-2.482164$   &                      &   $-2.465921$   &                           \\  
1.600   &   $-2.867620$    &   $-2.8669(5)^g$&  $-2.736687$     &  $-2.7330(8)^g$     &   $-2.698755$   &                      &   $-2.682563$   & $-2.6665(26)^g$           \\
2.000   &   $-2.999708$    &   $-2.9982^h$   &  $-2.867135$     &                     &   $-2.829137$   &                      &   $-2.813101$   &                           \\ 
5.000   &   $-3.768199$    &   $-3.7667^h$   &  $-3.624558$     &                     &   $-3.584967$   &                      &   $-3.568584$   &                           \\ 
10.000  &   $-4.627450$    &   $-4.626^h$    &  $-4.473459$     &                     &   $-4.432177$   &                      &   $-4.415350$   &                           \\
20.000  &   $-5.772448$    &   $-5.771^h$    &  $-5.607619$     &                     &   $-5.564582$   &                      &   $-5.547290$   &                           \\
50.000  &   $-7.815256$    &   $-7.814^h$    &  $-7.635847$     &                     &   $-7.590520$   &                      &   $-7.563695$   &                           \\
100.00  &   $-9.843074$    &   $-9.842^h$    &  $-9.652632$     &                     &   $-9.605634$   &                      &   $-9.586289$   &                           \\ \hline
\end{tabular}

\end{minipage}
\vspace{1mm}

$^a$ Baker et al. (1989) \cite{baker89}

$^b$ Drake et al. (1992) \cite{drake}

$^f$ Jones et al. (1996) \cite{ceperleyHF}

$^g$ Jones et al. (1997) \cite{ceperleyRPQMC}

$^h$ Ivanov (1994) \cite{ivanov94}

$^i$ Thurner et al. (1993) \cite{thurner}
}

\end{sidewaystable}

\begin{sidewaystable}
{\small
\begin{minipage}{18.5cm}
\caption{ Total energies $E$ of the states $\nu^10^-$, $1\le \nu \le 6$ as a function of the magnetic 
field strength $B$ as well as field free reference values and the values for finite field, if available. 
For $B>0.1$a.u., the energy values for the $6^10^-$ state are not optimally converged}
\vspace{1mm}

\begin{tabular}{|l|l|l|l|l|l|l|}\hline\hline
$B$     &  $ E(1^10^-)$    &  $ E(2^10^-)$   &  $ E(3^10^-)$  &  $ E(4^10^-)$     &   $ E(5^10^-) $   &  $ E(6^10^-) $      \\ \hline
        &                  &                 &                &                   &                   &                        \\
0.0000  &  $-2.123412$     &$-2.055015     $ &$-2.031252  $   &$-2.031014    $    &$-2.019999     $   &  $-2.019877 $          \\
        &$-2.123843430864^b$&$-2.0551463620^b$ &$-2.0312551443^b$&$-2.03106965^b$    &$-2.020002937^b$   &  $-2.019905990^b$   \\
0.0008  &  $-2.123413$     &$-2.055009     $ &$-2.031238  $   &$-2.030994    $    &$-2.019958     $   &  $-2.019824 $          \\    
0.004   &  $-2.123388$     &$-2.054868     $ &$-2.030983  $   &$-2.030463    $    &$-2.019427     $   &  $-2.018337 $          \\ 
0.008   &  $-2.123314$     &$-2.054438     $ &$-2.030349  $   &$-2.028842    $    &$-2.018198     $   &  $-2.014716 $          \\ 
0.020   &  $-2.122782$     &$-2.051758     $ &$-2.026727  $   &$-2.020802    $    &$-2.012781     $   &  $-2.005504 $          \\    
0.040   &  $-2.120940$     &$-2.044641     $ &$-2.017663  $   &$-2.006440    $    &$-2.000816     $   &  $-1.994681 $          \\    
0.080   &  $-2.114310$     &$-2.027515     $ &$-1.997336  $   &$-1.983478    $    &$-1.976059     $   &  $-1.970825 $          \\
0.160   &  $-2.093203$     &$-1.990787     $ &$-1.957118  $   &$-1.942289    $    &$-1.934541     $   &  \\
0.240   &  $-2.065812$     &$-1.952090     $ &$-1.916192  $   &$-1.900745    $    &$-1.892719     $   &  \\  
0.400   &  $-2.000350$     &$-1.870342     $ &$-1.831436  $   &$-1.815130    $    &$-1.806747     $   &  \\
0.500   &  $-1.954646$     &$-1.816921     $ &$-1.776662  $   &$-1.759971    $    &$-1.751519     $   &  \\   
        &  $-1.95455^e$    &                 &                &                   &                   &  \\   
0.800   &  $-1.802966$     &$-1.648143     $ &$-1.605029  $   &$-1.587531    $    &$-1.578773     $   &  \\
1.000   &  $-1.692794$     &$-1.529609     $ &$-1.485166  $   &$-1.467297    $    &$-1.458395     $   &  \\
        &  $-1.69185^e$    &                 &                &                   &                   &  \\   
1.600   &  $-1.331851$     &$-1.151043     $ &$-1.103929  $   &$-1.085324    $    &$-1.076152     $   &  \\
2.000   &  $-1.072194$     &$-0.883229     $ &$-0.834929  $   &$-0.816003    $    &$-0.806576     $   &  \\ 
        &  $-1.06905^e$    &                 &                &                   &                   &  \\   
5.000   &  $ 1.137045$     &$ 1.355762     $ &$ 1.408108  $   &$ 1.428117    $    &$ 1.437859     $   &  \\
10.000  &  $ 5.270657$     &$ 5.506354     $ &$ 5.560836  $   &$ 5.581403    $    &$ 5.591347     $   &  \\
20.000  &  $14.124717$     &$14.372781     $ &$14.428726  $   &$14.449668    $    &$14.459766     $   &  \\ 
50.000  &  $42.086569$     &$42.346227     $ &$42.403449  $   &$42.424712    $    &$42.434863     $   &  \\ 
100.000 &  $90.064304$     &$90.330942     $ &$90.388895  $   &$90.410339    $    &$90.420669     $   &  \\   \hline
\end{tabular}		
\end{minipage}

\vspace{1mm}

$^b$ Drake et al. (1992) \cite{drake}

$^e$ Larsen (1979) \cite{larsen}

}
\end{sidewaystable}

\begin{sidewaystable}
{\small
\begin{minipage}{23.7cm}
\caption{ Total energies $E$ of the triplet states $\nu^3_{-1}0^-$ ($S_z=-1$) for $1\le \nu\le 6$ as a 
function of the magnetic field strength $B$. We have also provided the results given in the literature so 
far.}
\vspace{1mm}

\begin{tabular}{|l||l      |l                ||l               |l                    ||l               |l                     ||l             |l                     ||l             |l                     ||l             |l                     | }\hline\hline
\rule[-3mm]{0mm}{8mm}
      & \multicolumn{2}{|c||}{ $E(1^30^-)$} 
      & \multicolumn{2}{c||}{ $E(2^30^-)$}   
      & \multicolumn{2}{c||}{ $E(3^30^-)$}   
      & \multicolumn{2}{c||}{ $E(4^30^-)$}   
      & \multicolumn{2}{c||}{ $E(5^30^-)$}   
      & \multicolumn{2}{c|}{ $E(6^30^-)$}    \\ \hline
\multicolumn{1}{|c||}{ $B$ }
      & this work  &   literature    & this work &   literature    & this work &   literature     & this work &  lt.  & this work &  lt.  & this work &   lt.       \\ \hline
      &            &                 &           &                 &           &                  &           &       &           &       &           & \\
0.0000& $-2.132910$&   $-2.133164$   &$-2.058016$&  $-2.058081$    &$-2.032298$& $-2.032324$      &$-2.031252$&       &$-2.020538$&       &           &           \\
      &            &$\;\;\;\;\;191^b$&           &$\;\;\;\;\;084^b$&           & $\;\;\;\;\;354^b$&\multicolumn{2}{l||}{$-2.031255168^b$}
												                     &\multicolumn{2}{l||}{$-2.020551187^b$}
																	 &\multicolumn{2}{l|}{$-2.020002957^b$}     \\
0.0008&$-2.133710 $&   $-2.1323^f$   &$-2.058810$&  $-2.0584^f$    &$-2.033081$&                  &$-2.032038$&       &$-2.021293$&       &$-2.020753$&  \\
0.004 &$-2.136889 $&   $-2.1344^f$   &$-2.061884$&  $-2.0615^f$    &$-2.035896$&                  &$-2.034870$&       &$-2.023839$&       &$-2.022647$&  \\
0.008 &$-2.140826 $&   $-2.1408(8)^g$&$-2.065499$&  $-2.0655(5)^g$ &$-2.039077$& $-2.0350(10)^g$  &$-2.037514$&       &$-2.026642$&       &$-2.023039$&  \\
0.020 &$-2.152378 $&                 &$-2.075054$&                 &$-2.047603$&                  &$-2.041411$&       &$-2.033564$&       &$-2.025840$&  \\
0.040 &$-2.170822 $&   $-2.169276^i$ &$-2.088326$&  $-2.086102^i$  &$-2.059120$&                  &$-2.046510$&       &$-2.041394$&       &$-2.035472$&  \\
0.080 &$-2.205130 $&   $-2.2050(6)^g$&$-2.111478$&  $-2.1100(3)^g$ &$-2.079242$& $-2.0578(14)^g$  &$-2.064505$&       &$-2.056648$&       &$-2.051347$&  \\
0.160 &$-2.266575 $&   $-2.262806^i$ &$-2.154912$&  $-2.112992^i$  &$-2.118891$&                  &$-2.103193$&       &$-2.095063$&       &&  \\
0.240 &$-2.322032 $&   $-2.3198^f$   &$-2.196517$&  $-2.1868^f$    &$-2.157959$&                  &$-2.141609$&       &$-2.133225$&       &&  \\
0.400 &$-2.422361 $&   $-2.4196^f$   &$-2.275521$&  $-2.2463^f$    &$-2.233362$&                  &$-2.216039$&       &$-2.207275$&       &&  \\
0.500 &$-2.480172 $&                 &$-2.322570$&                 &$-2.278711$&                  &$-2.260925$&       &$-2.252044$&       &&  \\ 
0.800 &$-2.638222 $&   $-2.6397(9)^g$&$-2.455054$&  $-2.4545(6)^g$ &$-2.407425$& $-2.3925(7)^g$   &$-2.388624$&       &$-2.379366$&       &&  \\
1.000 &$-2.733813 $&                 &$-2.537231$&                 &$-2.487761$&                  &$-2.468471$&       &$-2.459030$&       &&  \\  
1.600 &$-2.987185 $&   $-2.9873(8)^g$&$-2.760330$&  $-2.7596(8)^g$ &$-2.706990$& $-2.6749(6)^g$   &$-2.686687$&       &$-2.676877$&       &&  \\
2.000 &$-3.135142 $&                 &$-2.893357$&                 &$-2.838224$&                  &$-2.817461$&       &$-2.807405$&       &&  \\ 
5.000 &$-3.959235 $&                 &$-3.657879$&                 &$-3.596162$&                  &$-3.573739$&       &$-3.563137$&       &&  \\ 
10.000&$-4.848590 $&                 &$-4.509715$&                 &$-4.444111$&                  &$-4.420729$&       &$-4.409789$&       &&  \\
20.000&$-6.011488 $&                 &$-5.645191$&                 &$-5.576755$&                  &$-5.552682$&       &$-5.541487$&       &&  \\
50.000&$-8.059466 $&                 &$-7.672999$&                 &$-7.602384$&                  &$-7.577781$&       &$-7.566423$&       &&  \\
100.00&$-10.079973$&                 &$-9.688173$&                 &$-9.616899$&                  &$-9.592135$&       &$-9.580675$&       &&  \\ \hline
\end{tabular}

\end{minipage}
\vspace{1mm}

$^b$ Drake et al. (1992) \cite{drake}

$^f$ Jones et al. (1996) \cite{ceperleyHF}

$^g$ Jones et al. (1997) \cite{ceperleyRPQMC}

$^i$ Thurner et al. (1993) \cite{thurner}
}

\end{sidewaystable}

\begin{table}
\begin{minipage}{11.0cm}
\caption{Overview over all of the stationary points found in of the singlet transitions 
$\mu^10^+ \rightarrow \nu^10^-$ for $\mu = 1,...,5,\;\nu = 1,...,5$ in the range $0\le B \le 100a.u.$. 
The main part of the errors arises due to the interpolation over the relatively crude grid of field 
strengths. For high wavelengths, the finite accuracy of the energy values themselves contributes also to
error in the energies. }
\vspace{1mm}
\label{statsing}
\begin{tabular}{l   l   l   l}\hline\hline
           Component                        & wavelength/A         & position $B/a.u.$    & max/min  \\ 
$\nu^{2S+1}M^{(-1)^{\Pi_z}} $               &                      &                      &          \\ \hline
$ 2^10^+ \rightarrow 1^10^- $               & $  3665\pm 13      $ & $ 18.3\pm 0.3      $ &  min     \\	 
$ 2^10^+ \rightarrow 2^10^- $               & $  4291\pm 5       $ & $ 0.140\pm0.005    $ &  min     \\	 
$ 2^10^+ \rightarrow 2^10^- $               & $  5169\pm 11      $ & $ 0.899\pm0.003    $ &  max     \\	 
$ 2^10^+ \rightarrow 3^10^- $               & $  3258\pm 5       $ & $ 0.184\pm0.005    $ &  min     \\	 
$ 2^10^+ \rightarrow 3^10^- $               & $  3475\pm 6       $ & $ 0.68\pm0.02      $ &  max     \\	 
$ 2^10^+ \rightarrow 4^10^- $               & $  2949\pm 10      $ & $ 0.17\pm 0.02     $ &  min     \\	 
$ 2^10^+ \rightarrow 4^10^- $               & $  3073\pm 5       $ & $ 0.625\pm 0.005   $ &  max     \\	 
$ 2^10^+ \rightarrow 5^10^- $               & $  2797\pm 5       $ & $ 0.200\pm 0.005   $ &  min     \\	 
$ 2^10^+ \rightarrow 5^10^- $               & $  2905\pm 5       $ & $ 0.605\pm 0.005   $ &  max     \\	 
$ 3^10^+ \rightarrow 2^10^- $               & $ 90300\pm 3000    $ & $ 0.023\pm 0.001   $ &  max     \\
$ 3^10^+ \rightarrow 2^10^- $               & $ 43600\pm 1000    $ & $ 0.11\pm 0.02     $ &  min     \\
$ 3^10^+ \rightarrow 2^10^- $               & $ 18518\pm 100     $ & $ 8.9\pm 0.2       $ &  min     \\
$ 3^10^+ \rightarrow 3^10^- $               & $ 10655\pm 40      $ & $ 0.14\pm 0.01     $ &  min     \\
$ 3^10^+ \rightarrow 3^10^- $               & $ 17350\pm 40      $ & $ 1.45\pm 0.1      $ &  max     \\
$ 3^10^+ \rightarrow 4^10^- $               & $  7916\pm 40      $ & $ 0.18\pm 0.01     $ &  min     \\	 
$ 3^10^+ \rightarrow 4^10^- $               & $ 10250\pm 10      $ & $ 1.05\pm 0.10     $ &  max     \\	 
$ 3^10^+ \rightarrow 5^10^- $               & $  6969\pm 40      $ & $ 0.18\pm 0.01     $ &  min     \\
$ 3^10^+ \rightarrow 5^10^- $               & $  8538\pm 5       $ & $ 0.98\pm 0.06     $ &  max     \\
$ 4^10^+ \rightarrow 2^10^- $               & $  666000\pm 190000$ & $ 0.008\pm 0.004   $ &  min     \\
$ 4^10^+ \rightarrow 3^10^- $               & $  18000\pm 12     $ & $ 0.0188\pm 0.0002 $ &  min     \\
$ 4^10^+ \rightarrow 3^10^- $               & $  52800\pm 110    $ & $  6.6\pm0.3       $ &  min     \\
$ 4^10^+ \rightarrow 4^10^- $               & $  13850\pm 200    $ & $  0.032\pm 0.002  $ &  min     \\
$ 4^10^+ \rightarrow 4^10^- $               & $  41630\pm 220    $ & $  1.75 \pm 0.01   $ &  max     \\
$ 4^10^+ \rightarrow 5^10^- $               & $  11530\pm 100    $ & $  0.025\pm 0.002  $ &  min     \\
$ 4^10^+ \rightarrow 5^10^- $               & $  22750\pm 30     $ & $  1.28 \pm 0.02   $ &  max     \\
$ 4^10^+ \rightarrow 6^10^- $               & $   9730\pm 12     $ & $  0.0239\pm 0.0003$ &  min     \\
$ 5^10^+ \rightarrow 4^10^- $               & $  25000\pm 2000   $ & $  0.059 \pm 0.001 $ &  min     \\
$ 5^10^+ \rightarrow 5^10^- $               & $  34390\pm 40     $ & $  0.0085\pm 0.0015$ &  max     \\
$ 5^10^+ \rightarrow 5^10^- $               & $  23000\pm 600    $ & $  0.067\pm 0.005  $ &  min     \\\hline
\end{tabular}
\end{minipage}
\end{table}

\begin{table}
\begin{minipage}{11.0cm}
\caption{Same as Table \ref{statsing} but for the triplet transitions $\mu^30^+ \rightarrow \nu^30^-$ 
for $\mu = 1,...,4,\;\nu = 1,...,6$ }
\vspace{1mm}

\begin{tabular}{l   l   l   l}\hline\hline
           Component                        & wavelength/A        & position $B/a.u.$   &   max/min     \\ 
        $\nu^{2S+1}M^{(-1)^{\Pi_z}}$        &                     &                     &               \\ \hline
$ 1^30^+ \rightarrow 1^30^- $               & $  1869 \pm 5 $     & $ 40 \pm 2        $ & min           \\	 
$ 1^30^+ \rightarrow 2^30^- $               & $  3184 \pm 3 $     & $ 0.224 \pm 0.005 $ & min           \\	 
$ 1^30^+ \rightarrow 2^30^- $               & $  4289\pm 3.5$     & $ 2.25  \pm 0.02  $ & max           \\	 
$ 1^30^+ \rightarrow 3^30^- $               & $  2503 \pm 3 $     & $ 0.281 \pm 0.005 $ & min           \\	 
$ 1^30^+ \rightarrow 3^30^- $               & $  2837 \pm 2 $     & $ 1.51 \pm 0.01   $ & max           \\	 
$ 1^30^+ \rightarrow 4^30^- $               & $  2292 \pm 3 $     & $ 0.293 \pm 0.005 $ & min           \\	 
$ 1^30^+ \rightarrow 4^30^- $               & $  2522 \pm 1 $     & $ 1.39\pm 0.01    $ & max           \\	 
$ 1^30^+ \rightarrow 5^30^- $               & $  2197 \pm 5 $     & $ 0.299 \pm 0.005 $ & min           \\	 
$ 1^30^+ \rightarrow 5^30^- $               & $  2394 \pm 1 $     & $ 1.34\pm 0.02    $ & max           \\	 
$ 2^30^+ \rightarrow 2^30^- $               & $ 56500\pm 2100 $   & $ 0.05 \pm 0.01   $ & max           \\
$ 2^30^+ \rightarrow 2^30^- $               & $ 39900\pm 300  $   & $ 0.17 \pm 0.01   $ & min           \\
$ 2^30^+ \rightarrow 2^30^- $               & $ 12090\pm 25   $   & $ 26.6\pm0.3      $ & min           \\
$ 2^30^+ \rightarrow 3^30^- $               & $  9347 \pm 3   $   & $ 0.24  \pm 0.01  $ & min           \\
$ 2^30^+ \rightarrow 3^30^- $               & $ 16130\pm 45   $   & $ 3.65\pm 0.15    $ & max           \\
$ 2^30^+ \rightarrow 4^30^- $               & $  6990 \pm 10  $   & $ 0.26  \pm 0.01  $ & min           \\	 
$ 2^30^+ \rightarrow 4^30^- $               & $  9180 \pm 5   $   & $ 2.23  \pm 0.06  $ & max           \\	 
$ 2^30^+ \rightarrow 5^30^- $               & $  6186 \pm 10  $   & $ 0.27  \pm 0.01  $ & min           \\
$ 2^30^+ \rightarrow 5^30^- $               & $  7630 \pm 3.5 $   & $ 1.87\pm 0.05    $ & max           \\
$ 3^30^+ \rightarrow 3^30^- $               & $ 18000 \pm 2000$   & $ 0.026 \pm 0.003 $ & min           \\
$ 3^30^+ \rightarrow 3^30^- $               & $ 37428 \pm 80  $   & $ 20.5 \pm 0.5    $ & min           \\
$ 3^30^+ \rightarrow 4^30^- $               & $ 12400 \pm 1000$   & $ 0.039 \pm 0.003 $ & min           \\
$ 3^30^+ \rightarrow 4^30^- $               & $  22200\pm 100 $   & $  0.14 \pm 0.02  $ & max           \\
$ 3^30^+ \rightarrow 4^30^- $               & $  21650\pm 50  $   & $  0.25 \pm 0.02  $ & min           \\
$ 3^30^+ \rightarrow 4^30^- $               & $ 40600 \pm 40  $   & $ 4.5\pm 0.3      $ & max           \\
$ 3^30^+ \rightarrow 5^30^- $               & $ 10400 \pm 300 $   & $ 0.030 \pm 0.002 $ & min           \\
$ 3^30^+ \rightarrow 5^30^- $               & $ 15950 \pm 100 $   & $ 0.135 \pm 0.015 $ & max           \\
$ 3^30^+ \rightarrow 5^30^- $               & $ 15450 \pm 50  $   & $ 0.275 \pm 0.015 $ & min           \\
$ 3^30^+ \rightarrow 5^30^- $               & $ 21000 \pm 55  $   & $ 2.6   \pm 0.3   $ & max           \\
$ 3^30^+ \rightarrow 6^30^- $               & $  9400 \pm 100 $   & $ 0.032 \pm 0.002 $ & min           \\
$ 4^30^+ \rightarrow 3^30^- $               & $ 477000\pm 15000 $ & $ 0.0225\pm 0.0015$ & max           \\
$ 4^30^+ \rightarrow 3^30^- $               & $ 150000\pm 90000 $ & $ 0.08\pm 0.02    $ & min           \\
$ 4^30^+ \rightarrow 4^30^- $               & $ 96500 \pm 1000  $ & $0.0062\pm 0.0002 $ & max           \\
$ 4^30^+ \rightarrow 4^30^- $               & $ 20150 \pm  300  $ & $0.064   \pm 0.001$ & min           \\
$ 4^30^+ \rightarrow 4^30^- $               & $ 206000\pm 5000  $ & $0.17    \pm 0.03 $ & max           \\
$ 4^30^+ \rightarrow 4^30^- $               & $ 181000\pm12000  $ & $0.26    \pm 0.04 $ & min           \\
$ 4^30^+ \rightarrow 5^30^- $               & $  31400\pm 400   $ & $0.014   \pm 0.001$ & max           \\
$ 4^30^+ \rightarrow 5^30^- $               & $ 16800 \pm 500   $ & $0.068   \pm 0.005$ & min           \\
$ 4^30^+ \rightarrow 5^30^- $               & $ 44700 \pm3000   $ & $0.18    \pm 0.02 $ & max           \\
$ 4^30^+ \rightarrow 5^30^- $               & $ 38000 \pm6000   $ & $0.31    \pm 0.03 $ & min           \\\hline 
\end{tabular}
\end{minipage}
\end{table}

\newpage